\documentclass[graybox]{svmult}

\usepackage{mathptmx}       
\usepackage{helvet}         
\usepackage{courier}        
\usepackage{type1cm}        
\usepackage{graphicx}        
\usepackage{multicol}        
\usepackage[bottom]{footmisc}

\usepackage{amsmath,amssymb}
\usepackage{esint}
\usepackage{dsfont}
\usepackage{pst-all}
\usepackage{mathrsfs}

\newcommand{\R}{\mathbb{R}}
\newcommand{\N}{\mathbb{N}}
\newcommand{\Z}{\mathbb{Z}}
\newcommand{\C}{\mathbb{C}}
\newcommand{\cC}{\mathcal{C}}
\newcommand{\cK}{\mathcal{K}}
\newcommand{\cL}{\mathcal{L}}

\newcommand{\cQ}{\mathcal{Q}}
\newcommand{\cR}{\mathcal{R}}
\newcommand{\cS}{\mathcal{S}}

\newcommand{\gS}{\mathfrak{S}}
\newcommand{\dps}{\displaystyle }
\newcommand{\tr}{{\rm Tr}}
\renewcommand{\div}{{\rm div}}
\newcommand{\br}{\bold{r}}
\newcommand{\bR}{\bold{R}}
\newcommand{\bk}{\bold{k}}
\newcommand{\bq}{\bold{q}}
\newcommand{\bK}{\bold{K}}

\newcommand{\mnu}{m}
\newcommand{\Gper}{\gamma^0_{\rm per}}
\newcommand{\Hper}{H^0_{\rm per}}
\newcommand{\EF}{\epsilon^0_{\rm F}}
\newcommand{\ri}{i}
\DeclareMathAlphabet{\mathpzc}{OT1}{pzc}{m}{it}
\newcommand{\cFt}{\mathpzc{F}\!}

\begin{document}

\title*{The Microscopic Origin of the
Macroscopic Dielectric Permittivity of Crystals: A~Mathematical Viewpoint.}
\titlerunning{The Microscopic Origin of the
Macroscopic Dielectric Permittivity of Crystals}
\author{\'Eric Canc\`es, Mathieu Lewin and Gabriel Stoltz}
\institute{\'Eric Canc\`es and Gabriel Stoltz \at Universit\'e Paris-Est, CERMICS, Project-team Micmac, INRIA-Ecole des Ponts,  6 \& 8 avenue Blaise Pascal, 77455 Marne-la-Vall\'ee Cedex 2, France, \email{cances@cermics.enpc.fr, stoltz@cermics.enpc.fr}
\and Mathieu Lewin \at CNRS \& Laboratoire de Math\'ematiques UMR 8088, Universit\'e de Cergy-Pontoise, 95300 Cergy-Pontoise, France \email{Mathieu.Lewin@math.cnrs.fr}}
%
%
\maketitle

\abstract{The purpose of this paper is to provide a mathematical analysis of the Adler-Wiser formula relating the macroscopic relative permittivity tensor to the microscopic structure of the crystal at the atomic level. The technical level of the presentation is kept at its minimum to emphasize the mathematical structure of the results. We also briefly review some models describing the electronic structure of finite systems, focusing on density operator based formulations, as well as the Hartree model for perfect crystals or crystals with a defect.}

\section{Introduction}
\label{sec:introduction}

Insulating crystals are dielectric media. When an external electric field is applied, such an insulating material polarizes, and this induced polarization in turn affects the electric field. At the macroscopic level and in the time-independent setting, this phenomenon is modelled by the constitutive law
\begin{equation} \label{eq:DE}
D = \epsilon_0 \epsilon_{\rm M}  E
\end{equation}
specifying the relation between the macroscopic displacement field $D$ and the macroscopic electric field $E$. The constant $\epsilon_0$ is the dielectric permittivity of the vacuum, and $\epsilon_{\rm M}$ the macroscopic relative permittivity of the crystal, a $3 \times 3$ symmetric tensor such that $\epsilon_{\rm M} \ge 1$ in the sense of symmetric matrices ($\bk^T\epsilon_{\rm M}\bk \ge |\bk|^2$ for all $\bk \in \R^3$). This tensor is proportional to the identity matrix for isotropic crystals. Recall that $D$ is related to the so-called free charge $\rho_{\rm f}$ by the Gauss law $\div(D)=\rho_{\rm f}$ and that the macroscopic electric field $E$ is related to the macroscopic potential $V$ by $E=-\nabla V$, yielding the macroscopic Poisson equation
\begin{equation} \label{eq:macroscopicPoisson}
-\div (\epsilon_M \nabla V) = \rho_{\rm f}/{\epsilon_0}.
\end{equation}

\medskip

In the time-dependent setting, (\ref{eq:DE}) becomes a time-convolution product:
\begin{equation} \label{eq:TDDE}
D(\br,t) = \epsilon_0 \int_{-\infty}^{+\infty} \epsilon_{\rm M}(t-t') E(\br,t') \, dt'.
\end{equation}
Fourier transforming in time, we obtain
$$
\cFt D(\br,\omega) =  \cFt \epsilon_{\rm M}(\omega) \cFt E(\br,\omega),
$$
where, as usual in Physics, we have used the following normalization convention for the Fourier transform with respect to the time-variable: 
$$
\cFt f(\br,\omega) = \int_{-\infty}^{+\infty} f(\br,t) \, \mathrm{e}^{i\omega t} \, dt
$$
(note that there is no minus sign in the phase factor). The time-dependent tensor $\epsilon_{\rm M}$ in (\ref{eq:DE}) can be seen as the zero-frequency limit of the frequency-dependent tensor $\cFt\epsilon_{\rm M}(\omega)$. 

Of course, the constitutive laws (\ref{eq:DE}) (time-independent case) and (\ref{eq:TDDE}) (time-dependent case) are only valid in the \emph{linear response regime}. When strong dielectric field are applied, the response can be strongly nonlinear. 

\medskip

The purpose of this paper is to provide a mathematical analysis of the Adler-Wiser formula~\cite{Adler-62,Wiser-63} relating the macroscopic relative permittivity tensor $\epsilon_{\rm M}$ (as well as the frequency-dependent tensor $\cFt\epsilon_{\rm M}(\omega)$) to the microscopic structure of the crystal at the atomic level.

In Section~\ref{sec:Hartree}, we discuss the modelling of the electronic structure of finite molecular systems. We introduce in particular the Hartree model (also called reduced Hartree-Fock model in the mathematical literature), which is the basis for our analysis of the electronic structure of crystals. This model is an approximation of the electronic $N$-body Schr\"odinger equation allowing to compute the ground state electronic density of a molecular system containing $M$ nuclei considered as classical particles (Born-Oppenheimer approximation) and $N$ quantum electrons, subjected to Coulomb interactions. The only empirical parameters in this model are a few fundamental constants of Physics (the reduced Planck constant $\hbar$, the mass of the electron $m_{\rm e}$, the elementary charge $e$, and the dielectric permittivity of the vacuum $\epsilon_0$) and the masses and charges of the nuclei. In this respect, this is an {\it ab initio}, or first-principle, model in the sense that it does not contain any empirical parameter specific to the molecular system under consideration. 

We then show, in Section~\ref{sec:crystals}, how to extend the Hartree model for molecular systems (finite number of particles) to crystals (infinite number of particles). We first deal with perfect crystals (Section~\ref{sec:perfect}), then with crystals with local defects (Section~\ref{sec:RHFperturbed}). The mathematical theory of the electronic structure of crystals with local defects presented here (and originally published in \cite{CanDelLew-08a}) has been strongly inspired by previous works on the mathematical foundations of quantum electrodynamics (QED) \cite{HaiLewSer-05a,HaiLewSol-07,HaiLewSerSol-07}. In some sense, a defect embedded in a insulating or semi-conducting crystal behaves similarly as a nucleus embedded in the polarizable vacuum of QED.

In Section~\ref{sec:dielectric}, we study the dielectric response of a crystal. First, we focus on the response to an effective time-independent potential $V$, and expand it in powers of $V$ (Section~\ref{sec:expansion}). The linear response term allows us to define the (microscopic) dielectric operator $\epsilon$ and its inverse $\epsilon^{-1}$, the (microscopic) dielectric permittivity operator, and also to define a notion of renormalized charge for defects in crystals (Section~\ref{sec:Qsmall}). In Section~\ref{sec:macro}, we derive the Adler-Wiser formula from the Hartree model, by means of homogenization arguments. Loosely speaking, a defect in a crystal generates an external field and thereby a dielectric response of the crystal. If a given local defect is properly rescaled, it produces a macroscopic charge (corresponding to the free charge $\rho_{\rm f}$ in (\ref{eq:macroscopicPoisson}) and the total Coulomb potential converges to the macroscopic potential $V$ solution to  (\ref{eq:macroscopicPoisson}) where $\epsilon_{\rm M}$ is the tensor provided by the Adler-Wiser formula. A similar strategy can be used to obtain the frequency-dependent tensor $\cFt\epsilon_{\rm M}(\omega)$ (Section~\ref{sec:TDexpansion}). 

As trace-class and Hilbert-Schmidt operators play a central role in the mathematical theory of electronic structure, their definitions and some of their basic properties are recalled in Appendix for the reader's convenience. 

\medskip

The mathematical results contained in this proceeding have been published \cite{CanDelLew-08a,CanDelLew-08b,CanLew10}, or will be published very soon~\cite{CanSto}. The proofs are omitted. A pedagogical effort has been made to present this difficult material to non-specialists.

\medskip

As usual in first-principle modelling, we adopt the system of atomic units, obtained by setting
$$
\hbar =1, \quad m_{\rm e}=1, \quad e=1, \quad \frac{1}{4\pi\epsilon_0}=1,
$$
so that (\ref{eq:macroscopicPoisson2}) reads in this new system of units:
\begin{equation} \label{eq:macroscopicPoisson2}
-\div (\epsilon_M \nabla V) = 4\pi \rho_{\rm f}.
\end{equation}
For simplicity, we omit the spin variable, but taking the spin into account does not add any difficulty. It simply makes the mathematical formalism a little heavier.

\section{Electronic structure models for finite systems}
\label{sec:Hartree}

Let ${\cal H}$ be a Hilbert space and $\langle \cdot | \cdot \rangle$ its inner product (bra-ket Dirac's notation). Recall that if $A$ is a self-adjoint operator on ${\cal H}$ and $\phi$ and $\psi$ are in $D(A)$, the domain of~$A$, then $\langle \phi|A|\psi \rangle := \langle \phi | A\psi \rangle = \langle A \phi | \psi \rangle$. If $A$ is bounded from below, the bilinear form $(\phi,\psi) \mapsto  \langle \phi | A|\psi \rangle$ can be extended in a unique way to the form domain of $A$. For instance, the operator $A=-\Delta$ with domain $D(A)=H^2(\R^d)$ is self-adjoint on $L^2(\R^d)$. Its form domain is $H^1(\R^d)$ and $\langle \phi | A|\psi \rangle = \int_{\R^d} \nabla \phi \cdot \nabla \psi$. In the sequel, we denote by ${\cal S}({\cal H})$ the vector space of {\em bounded} self-adjoint operators on ${\cal H}$.  

\medskip

For $k=0$, $1$ and $2$, and with the convention $H^0(\R^3)=L^2(\R^3)$, we denote by 
$$
\bigwedge_{i=1}^N H^k(\R^3) := \left\{ \left. \Psi \in H^k(\R^{3N}) \; \right| \; \Psi(\br_{p(1)}, \cdots , \br_{p(N)}) = \epsilon(p) \Psi(\br_1, \cdots, \br_N), \; \forall p \in {\mathbb S}_N \right\}
$$
(where ${\mathbb S}_N$ is the group of the permutations of $\left\{1,\cdots,N\right\}$ and $\epsilon(p)$ the parity of $p$) the antisymmetrized tensor product of $N$ spaces $H^k(\R^3)$. These spaces are used to describe the electronic state of an $N$ electron system. The antisymmetric constraint originates from the fact that electrons are fermions.

\subsection{The $N$-body Schr\"odinger model}

Consider a molecular system with $M$ nuclei of charges $z_1, \cdots, z_M$. As we work in atomic units, $z_k$ is a positive integer. Within the Born-Oppenheimer approximation, the nuclei are modelled as classical point-like particles. This approximation results from a combination of an adiabatic limit (the small parameter being the square root of the ratio between the mass of the electron and the mass of the lightest nucleus present in the system), and a semi-classical limit. We refer to \cite{Friesecke1,Friesecke2} and references therein for the mathematical aspects. 

Usually, nuclei are represented by point-like particles. If the $M$ nuclei are located at points $\bR_1, \cdots, \bR_M$ of $\R^3$, the nuclear charge distribution is modelled by
$$
\rho^{\rm nuc} = \sum_{k=1}^M z_k \delta_{\bR_k},
$$
where $\delta_{\bR_K}$ is the Dirac measure at point $\bR_k$. The Coulomb potential generated by the nuclei and seen by the electrons then reads 
$$
V^{\rm nuc}(\br):=- \sum_{k=1}^M \frac{z_k}{|\br-\bR_k|}
$$
(the minus sign comes from the fact that the interaction between nuclei and electrons is attractive).
In order to avoid some technical difficulties due to the singularity of the potential generated by point-like nuclei, the latter are sometimes replaced with smeared nuclei:
$$
\rho^{\rm nuc}(\br) = \sum_{k=1}^M z_k \chi(\br-\bR_K),
$$
where $\chi$ is a smooth approximation of the Dirac measure $\delta_0$, or more precisely a non-negative smooth radial function such that $\int_{\R^3}\chi=1$, supported in a small ball centered at $0$. In this case, 
$$
V^{\rm nuc}(\br):=-(\rho^{\rm nuc} \star |\cdot|^{-1})(\br) = -\int_{\R^3} \frac{\rho^{\rm nuc}(\br')}{|\br-\br'|} \, d\br'
$$
is a smooth function. We will sometimes denote this smooth function by $V_{\rho^{\rm nuc}}$
in order to emphasize that the potential is generated by a non-singular charge distribution.

\medskip

The main quantity of interest in our study is the electrostatic potential generated by the total charge, which is by definition the sum of nuclear charge $\rho^{\rm nuc}$ and the electronic charge $\rho^{\rm el}$. According to the Born-Oppenheimer approximation, electrons are in their ground state, and $\rho^{\rm el}$ is a density associated with the ground state wavefunction $\Psi_0$. Let us make this definition more precise.

Any (pure) state of a system of $N$ electrons is entirely described by a wavefunction $\Psi \in  \bigwedge_{i=1}^N L^2(\R^3)$ satisfying the normalization condition $\|\Psi\|_{L^2(\R^{3N})}=1$. The density associated with $\Psi$ is the function $\rho_\Psi$ defined by
\begin{equation}\label{eq:defrhoel}
\rho_\Psi(\br) = N \int_{\R^{3(N-1)}} |\Psi(\br,\br_2,\cdots,\br_N)|^2 \, d\br_2 \cdots d\br_N.
\end{equation}
Clearly, 
$$
\rho_\Psi \ge 0, \quad \rho_\Psi \in L^1(\R^3), \quad \mbox{and} \quad \int_{\R^3}\rho_\Psi = N.
$$
It can be checked that if $\Psi \in \bigwedge_{i=1}^N H^1(\R^3)$, then $\sqrt{\rho} \in H^1(\R^3)$, which implies in particular that $\rho_\Psi \in L^1(\R^3) \cap L^3(\R^3)$.

The ground state wavefunction $\Psi_0$ is the lowest energy, normalized eigenfunction of the time-independent Schr\"odinger equation
\begin{equation} \label{eq:Schrodinger}
H_N \Psi = E \Psi, \quad \Psi \in \bigwedge_{i=1}^N H^2(\R^3), \quad \|\Psi\|_{L^2(\R^{3N})}=1,
\end{equation}
where $H_N$ is the electronic Hamiltonian. The latter operator is self-adjoint 
on $\bigwedge_{i=1}^N L^2(\R^3)$, with domain $\bigwedge_{i=1}^N H^2(\R^3)$ and form domain $\bigwedge_{i=1}^N H^1(\R^3)$, and is defined as
\begin{equation} \label{eq:HN}
H_N = -  \frac{1}{2}\sum_{i=1}^N \Delta_{\br_i}+ \sum_{i=1}^N V^{\rm nuc}(\br_i)+ \sum_{1 \le i < j \le N} \frac{1}{|\br_i-\br_j|}.
\end{equation}
The first term in the right-hand side of (\ref{eq:HN}) models the kinetic energy of the electrons, the second term the Coulomb interaction between nuclei and electrons and the third term the Coulomb interaction between electrons. For later purposes, we write
$$
H_N = T + V_{\rm ne} + V_{\rm ee},
$$
where
$$ 
T = -  \frac{1}{2}\sum_{i=1}^N \Delta_{\br_i}, \quad V_{\rm ne}= \sum_{i=1}^N V^{\rm nuc}(\br_i), \quad V_{\rm ee} = \sum_{1 \le i < j \le N} \frac{1}{|\br_i-\br_j|}.
$$
It is proved in \cite{Zhislin} that if the molecular system is neutral ($\sum_{k=1}^M z_k = N$) or positively charged ($\sum_{k=1}^M z_k \ge N$), then the essential spectrum of $H_N$ is an interval of the form $[\Sigma_N,+\infty)$ with $\Sigma_N\le 0$ and $\Sigma_N < 0$ if $N \ge 2$, and its discrete spectrum is an increasing infinite sequence of negative eigenvalues converging to $\Sigma_N$. This guarantees the existence of $\Psi_0$. If $E_0$, the lowest eigenvalue of $H_N$ is non-degenerate, $\Psi_0$ is unique up to a global phase, and $\rho^{\rm el}=\rho_{\Psi^0}$ is therefore uniquely defined by (\ref{eq:defrhoel}). If $E_0$ is degenerate, then the ground state electronic density is not unique. As the usual Born-Oppenheimer approximation is no longer valid when $E_0$ is degenerate, we will assume from now on that $E_0$ is a simple eigenvalue.

Note that $\Psi_0$ can also be defined variationally: It is the minimizer of 
\begin{equation} \label{eq:varPsi0}
\inf \left\{ \langle \Psi|H_N|\Psi \rangle, \; \Psi \in \bigwedge_{i=1}^N H^1(\R^3), \; \|\Psi\|_{L^2(\R^{3N})}=1 \right\}.
\end{equation}
Otherwise stated, it is obtained by minimizing the energy $\langle \Psi|H_N|\Psi \rangle$ over the set of all normalized, antisymmetric wavefunctions $\Psi$ of finite energy.

\medskip

Let us mention that, as in the absence of magnetic field, the $N$-body Hamiltonian is real (in the sense that it transforms a real-valued function into a real-valued function), there is no loss of generality in working in the space of real-valued $N$-body wavefunctions. Under the assumption that $E_0$ is non-degenerate, (\ref{eq:varPsi0}) has exactly two minimizers, $\Psi_0$ and $-\Psi_0$, both of them giving rise to the same electronic density.

\subsection{The $N$-body Schr\"odinger model for non-interacting electrons}
\label{sec:noninteraction}

Neither the Schr\"odinger equation (\ref{eq:Schrodinger}) nor the minimization (\ref{eq:varPsi0}) can be solved with standard numerical techniques when $N$ exceeds two or three. On the other hand, these problems become pretty simple when the interaction between electrons is neglected. In this case, the $N$-body Hamiltonian is separable and reads 
$$
H_N^0 = T+V_{\rm ne}=\sum_{i=1}^N h_{\br_i} \quad 
\mbox{where} \quad h_{\br_i} = - \frac 12 \Delta_{\br_i} + V^{\rm nuc}
$$
is a self-adjoint operator on $L^2(\R^3)$ with domain $H^2(\R^3)$ and form domain $H^1(\R^3)$, acting on functions of the variable~$\br_i$. It is known that the essential spectrum of $h$ is $[0,+\infty)$ and that the discrete spectrum of $h$ is an increasing infinite sequence of negative eigenvalues converging to $0$. Let us denote by $\epsilon_1 < \epsilon_2 \le \epsilon_3 \le \cdots$ the eigenvalues of $h$ counted with their multiplicities (it can be shown that $\epsilon_1$ is simple) and let $(\phi_i)_{i \ge 0}$ be an orthonormal family of associated eigenvectors:
$$
h\phi_i = \epsilon_i\phi_i, \quad \epsilon_1 < \epsilon_2 \le \epsilon_3 \le \cdots, \quad \phi_i \in H^2(\R^3), \quad \langle \phi_i|\phi_j\rangle_{L^2(\R^3)} = \delta_{ij}.
$$
The eigenfunctions $\phi_i$ are called (molecular) orbitals and the eigenvalues $\epsilon_i$ are called (one-particle) energy levels. 

\medskip

It is easy to check that if $\epsilon_{N} < \epsilon_{N+1}$, then  
\begin{equation} \label{eq:varPsi00}
\inf \left\{ \langle \Psi|H_N^0|\Psi \rangle, \; \Psi \in \bigwedge_{i=1}^N H^1(\R^3), \; \|\Psi\|_{L^2(\R^{3N})}=1 \right\}
\end{equation}
has a unique solution (up to a global phase) given by the Slater determinant
\begin{equation} \label{eq:Slater}
\Psi_0(\br_1, \cdots, \br_N) = \frac{1}{\sqrt{N!}} \left| \begin{array}{cccccc}
\phi_1(\br_1) & \phi_1(\br_2) & \cdot & \cdot & \cdot & \phi_1(\br_N) \\
\phi_2(\br_1) & \phi_2(\br_2) & \cdot & \cdot &  \cdot & \phi_2(\br_N) \\
\cdot & \cdot & \cdot & \cdot & \cdot & \cdot  \\
\cdot & \cdot & \cdot & \cdot & \cdot & \cdot  \\
\cdot & \cdot & \cdot & \cdot & \cdot & \cdot  \\
\phi_N(\br_1) & \phi_N(\br_2) & \cdot & \cdot &  \cdot & \phi_N(\br_N) 
 \end{array} \right|,
\end{equation}
and that the ground state electronic density (\ref{eq:defrhoel}) takes the simple form$$
\rho^{\rm el}(\br) = \sum_{i=1}^N |\phi_i(\br)|^2.
$$ 

The above description of the electronic states of a set of $N$ non-interacting electrons in terms of orbitals cannot be easily extended to infinite systems such as crystals (the number of orbitals becoming infinite). For this reason, we introduce a new formulation based on the concept of one-particle density operator, here abbreviated as density operator.

\subsection{Density operators}

The (one-particle) density operator of a system of $N$ electrons is an element of the convex set
$$
{\cal D}_N=\left\{ \gamma \in {\cal S}(L^2(\R^3)) \; | \; 0 \le \gamma \le 1, \; \tr(\gamma) = N \right\}.
$$
Recall that if $A$ and $B$ are two bounded self-adjoint operators on a Hilbert space ${\cal H}$, the notation $A \le B$ means that $\langle \psi|A|\psi\rangle \le \langle \psi|B|\psi\rangle$ for all $\psi \in {\cal H}$. 

Any density operator $\gamma \in {\cal D}_N$ is trace-class, hence compact (the basic properties of trace-class operators are recalled in the Appendix). It can therefore be diagonalized in an orthonormal basis:
\begin{equation} \label{eq:dec_gamma}
\gamma = \sum_{i=1}^{+\infty} n_i |\phi_i\rangle \langle \phi_i| \quad \mbox{with} \quad \langle \phi_i|\phi_j\rangle = \delta_{ij}.
\end{equation}
The eigenvalues $n_i$ are called occupation numbers; the eigenfunctions $\phi_i$ are called natural orbitals. The conditions $0 \le \gamma \le 1$ and $\tr(\gamma)=N$ are respectively equivalent to 
$$
0 \le n_i \le 1 \quad \mbox{and} \qquad \sum_{i=1}^{+\infty} n_i = N.
$$
The fact that $0 \le n_i \le 1$ is a mathematical translation of the Pauli exclusion principle, stipulating that each quantum state $|\phi_i\rangle$ is occupied by at most one electron. The sum of the occupation numbers is equal to $N$, the number of electrons in the system. The density associated with $\gamma$ is defined by
\begin{equation} \label{eq:rhogamma}
\rho_\gamma(\br)= \sum_{i=1}^{+\infty} n_i |\phi_i(\br)|^2,
\end{equation}
this definition being independent of the choice of the orthonormal basis $(\phi_i)_{i \ge 1}$ in~(\ref{eq:dec_gamma}) and satisfies
$$
\rho_\gamma \ge 0, \quad \rho_\gamma \in L^1(\R^3), \quad \mbox{and} \quad \int_{\R^3} \rho_\gamma = N.
$$
The kinetic energy of the density operator $\gamma$ is defined as 
$$
T(\gamma):=\frac 12 \tr(|\nabla|\gamma|\nabla|),
$$
and can be finite or infinite. Recall that the operator $|\nabla|$ is the unbounded self-adjoint operator on $L^2(\R^3)$ with domain $H^1(\R^3)$ defined by
$$
\forall \phi \in H^1(\R^3), \quad ({\cal F}(|\nabla|\phi))(\bk) = |\bk| ({\cal F}(\phi))(\bk)
$$
where ${\cal F}$ is the unitary Fourier transform
\[
{\cal F} \phi(\bk) = \widehat{\phi}(\bk) 
= \frac{1}{(2\pi)^{3/2}} \int_{\R^3} \phi(\br) \, \mathrm{e}^{-i \bk \cdot \br} \, d\br.
\]
The kinetic energy of a density operator $\gamma$ decomposed as (\ref{eq:dec_gamma}) is finite if and only if each $\phi_i$ is in $H^1(\R^3)$ and $\sum_{i=1}^{+\infty}n_i\|\nabla\phi_i\|_{L^2(\R^3)}^2 < \infty$, in which case
$$
T(\gamma)= \frac 12 \sum_{i=1}^{+\infty} n_i\|\nabla\phi_i\|_{L^2(\R^3)}^2.
$$
As $|\nabla|$ is the square root of $-\Delta$ (\textit{i.e.} $|\nabla|$ is self-adjoint, positive and $|\nabla|^2=-\Delta$), the element $\tr(|\nabla|\gamma|\nabla|)$ of $\R_+ \cup \{+\infty\}$ is often denoted by $\tr(-\Delta\gamma)$. Using this notation, we can define the convex set ${\cal P}_N$ of the density operators of finite energy as 
$$
{\cal P}_N= \left\{ \gamma \in {\cal S}(L^2(\R^3)) \; | \; 0 \le \gamma \le 1, \; \tr(\gamma) = N, \, \tr(-\Delta\gamma) < \infty \right\}.
$$
Lastly, it is sometimes useful to introduce the integral kernel of a density operator $\gamma \in {\cal P}_N$, which is called a (one-particle) density matrix, and is usually also denoted by $\gamma$. It is by definition the function $\gamma \in L^2(\R^3\times\R^3)$ such that
\begin{equation} \label{eq:defgamma2}
\forall \phi \in L^2(\R^3), \quad (\gamma\phi)(\br) = \int_{\R^3} \gamma(\br,\br') \phi(\br') \, d\br'.
\end{equation}
The expression of the density matrix $\gamma$ in terms of natural orbitals and occupation numbers thus reads
$$
\gamma(\br,\br') = \sum_{i=1}^{+\infty} n_i \phi_i(\br) \, \phi_i(\br').
$$ 
Formally $\rho_\gamma(\br)=\gamma(\br,\br)$ and this relation makes sense rigorously as soon as the density matrix $\gamma$ has a trace on the three-dimensional vector subspace $\left\{(\br,\br), \, \br \in \R^3 \right\}$ of $\R^3\times\R^3$.  

\medskip

Let us now clarify the link between the description of electronic structures in terms of wavefunctions and the one in terms of density operators.

The density matrix associated with a wavefunction $\Psi \in \wedge_{i=1}^N L^2(\R^3)$ such that $\|\Psi\|_{L^2(\R^{3N})}=1$ is the function of $L^2(\R^3 \times \R^3)$ defined as
\begin{equation} \label{eq:defgamma1}
\gamma_\Psi(\br,\br') = N \int_{\R^{3(N-1)}} \Psi(\br,\br_2,\cdots,\br_N) \Psi(\br',\br_2,\cdots,\br_N) \, d\br_2 \cdots d\br_N
\end{equation}
(recall that we are dealing with real-valued wavefunctions), and the corresponding density operator by
\begin{equation} \label{eq:defgamma3}
\forall \phi \in L^2(\R^3), \quad (\gamma_\Psi\phi)(\br) = \int_{\R^3} \gamma_\Psi(\br,\br') \phi(\br') \, d\br'.
\end{equation}
It is easy to see that the density operator $\gamma_\Psi$ is in ${\cal D}_N$. Under the additional assumption that $\Psi \in \wedge_{i=1}^N H^1(\R^3)$, it is even in ${\cal P}_N$. Besides, the definition (\ref{eq:defrhoel}) of the density associated with $\Psi$ agrees with the definition (\ref{eq:rhogamma}) of the density associated with $\gamma_\Psi$, \textit{i.e.} 
$$
\rho_\Psi = \rho_{\gamma_\Psi},
$$ 
and the same holds with the definition of the kinetic energy if $\Psi \in \wedge_{i=1}^N H^1(\R^3)$:
$$
\langle \Psi |T| \Psi \rangle = T(\gamma_\Psi). 
$$

\medskip

\begin{remark} The maps $\left\{ \left. \Psi \in \bigwedge_{i=1}^N L^2(\R^3) \; \right| \; | \|\Psi\|_{L^2(\R^{3N})} = 1 \right\} \ni \Psi \mapsto \gamma_\Psi \in {\cal D}_N$ and $\left\{ \left. \Psi \in \bigwedge_{i=1}^N H^2(\R^3) \; \right| \; | \|\Psi\|_{L^2(\R^{3N})} = 1 \right\} \ni \Psi \mapsto \gamma_\Psi \in {\cal P}_N$ are not surjective. This means that an element of ${\cal D}_N$ (resp. of ${\cal P}_N$) is not necessarily the density operator associated with some {\em pure} state. However any $\gamma \in {\cal D}_N$ (resp. any $\gamma \in {\cal D}_N$) is the (one-particle) density operator associated with some {\em mixed} state (represented by some $N$-particle density operator). This property is referred to as the $N$-representability property of density operators.
\end{remark}

We can now reformulate the electronic structure problem {\em for a system of $N$ non-interacting electrons}, in terms of density operators:
\begin{enumerate}
\item The energy of a wavefunction $\Psi \in \wedge_{i=1}^N H^1(\R^3)$ is a linear form with respect to the density operator $\gamma_\Psi$:
$$
\langle \Psi | H^0_N | \Psi \rangle = E^0_{\rho^{\rm nuc}}(\gamma_\Psi) \quad \mbox{where} \quad E^0_{\rho^{\rm nuc}}(\gamma) = \tr\left( -\frac 12 \Delta \gamma \right) + \int_{\R^3} \rho_{\gamma}V^{\rm nuc} ;
$$
\item The ground state density matrix, that is the density operator associated with the ground state wavefunction $\Psi^0$ defined by (\ref{eq:varPsi00}), is the orthogonal projector (for the $L^2$ inner product) on the space $\mbox{Span}(\phi_1,\cdots,\phi_N)$:
$$
\gamma_{\Psi^0} = \sum_{i=1}^N |\phi_i\rangle \, \langle\phi_i|;
$$ 
\item The ground state energy and the ground state density operators are obtained by solving the minimization problem
\begin{equation}
\inf \left\{ E^0_{\rho^{\rm nuc}}(\gamma), \; \gamma \in {\cal S}(L^2(\R^3)), \; 0 \le \gamma \le 1, \; \tr(\gamma)=N, \, \tr(-\Delta \gamma) < \infty \right\}. \label{eq:minrelax}
\end{equation}
\end{enumerate}
The advantages of the density operator formulation, which are not obvious for finite systems, will clearly appear in Section~\ref{sec:crystals}, where we deal with crystals.

\subsection{The Hartree model and other density operator models of electronic structures}

Let us now reintroduce the Coulomb interaction between electrons, taking as a starting point the non-interacting system introduced in Section~\ref{sec:noninteraction}. The models presented in this section are density operator models in the sense that the ground state energy and density are obtained by minimizing some {\em explicit} functional $E_{\rho^{\rm nuc}}(\gamma)$ over the set of $N$-representable density operators ${\cal P}_N$.

\medskip

All these models share the same mathematical structure. They read:
\begin{equation} \label{eq:minmod}
\inf \left\{ E_{\rho^{\rm nuc}}(\gamma), \; \gamma \in {\cal S}(L^2(\R^3)), \; 0 \le \gamma \le 1, \; \tr(\gamma) = N, \; \tr(-\Delta \gamma) < \infty \right\},
\end{equation}
with 
$$
E_{\rho^{\rm nuc}}(\gamma) = \tr \left( -\frac 12 \Delta \gamma \right) + \int_{\R^3} \rho_\gamma V_{\rho^{\rm nuc}}  + \frac 12 D(\rho_\gamma,\rho_\gamma) + \widetilde E(\gamma),
$$
where
\begin{equation} \label{eq:defCoulomb}
D(f,g) = \int_{\R^3} \int_{\R^3} \frac{f(\br)\, g(\br')}{|\br-\br'|} \, d\br \, d\br'
\end{equation}
is the classical Coulomb interaction and $\widetilde E(\gamma)$ some correction term. Note that $D(f,g)$ is well defined for $f$ and $g$ in $L^{6/5}(\R^3)$, see for instance \cite[Section~IX.4]{ReeSim2}. Recall also that for each $\gamma \in {\cal P}_N$, $\rho_\gamma \in L^1(\R^3) \cap L^3(\R^3) \hookrightarrow L^{6/5}(\R^3)$. 

\medskip

The Hartree model, on which we will focus in this proceeding, corresponds to $\widetilde E(\gamma)=0$:
$$
E_{\rho^{\rm nuc}}^{\rm Hartree}(\gamma) = \tr \left( -\frac 12 \Delta \gamma \right) + \int_{\R^3}  \rho_\gamma V_{\rho^{\rm nuc}} + \frac 12 D(\rho_\gamma,\rho_\gamma).
$$
The reason why we study this model is that it has much nicer mathematical properties than
other models with $\widetilde E(\gamma) \not = 0$ (see below).

\medskip

The Kohn-Sham models \cite{KS} originate from the Density Functional Theory (DFT) \cite{DreGro-90}. In this kind of models, $\widetilde E(\gamma)$ is an explicit functional of the density $\rho_\gamma$, called the exchange-correlation functional:
\begin{equation} \label{eq:KS}
E_{\rho^{\rm nuc}}^{\rm KS}(\gamma) = \tr \left( -\frac 12 \Delta \gamma \right) + \int_{\R^3} V_{\rho^{\rm nuc}} \rho_\gamma + \frac 12 D(\rho_\gamma,\rho_\gamma) + E^{\rm xc}(\rho_\gamma).
\end{equation}
If follows from the Hohenberg-Kohn theorem \cite{HK} (see \cite{LiebDFT} for a more mathematical presentation of this result) that there exists some functional $E^{\rm xc}(\rho)$ depending only on the density $\rho$, such that minimizing (\ref{eq:minmod}) with $E_{\rho^{\rm nuc}}=E_{\rho^{\rm nuc}}^{\rm KS}$ provides the {\em exact} ground state energy and density, whatever the nuclear charge distribution $\rho^{\rm nuc}$. Note however, that the Kohn-Sham ground state density operator obtained by minimizing  (\ref{eq:minmod}) is not the ground state density operator corresponding to the ground state wavefunction $\Psi^0$. Unfortunately, the exact exchange-correlation functional is not known. Many approximate functionals have been proposed, and new ones come up on a regular basis. For the sake of illustration, the simplest approximate exchange-correlation functional (but clearly not the best one) is the so-called X$\alpha$ functional
$$
E^{\rm xc}_{{\rm X}\alpha}(\rho) = -C_{{\rm X}\alpha} \int_{\R^3} \rho^{4/3}, 
$$
where $C_{{\rm X}\alpha}$ is a positive constant

\medskip

Lastly, the models issued from the Density-Matrix Functional Theory (DMFT) involve functionals $\widetilde E(\gamma)$ depending explicitly on the density operator $\gamma$, but not only on the density $\rho_\gamma$. Similar to DFT, there exists an {\em exact} (but unknown) functional $\widetilde E(\gamma)$ for which minimizing (\ref{eq:minmod}) gives the exact ground state energy and density, whatever the nuclear charge distribution $\rho^{\rm nuc}$. However, unlike the exact DFT functional, the exact DMFT functional also provides the exact ground state density operator. Several approximate DMFT functionals have been proposed. Note that the Hartree-Fock model, which is usually defined as the variational approximation of (\ref{eq:varPsi0}) obtained by restricting the minimization set to the set of finite energy Slater determinants, can also be seen as a DMFT functional 
$$
E_{\rho^{\rm nuc}}^{\rm HF}(\gamma) = \tr \left( -\frac 12 \Delta \gamma \right) + \int_{\R^3}  \rho_\gamma V_{\rho^{\rm nuc}} + \frac 12 D(\rho_\gamma,\rho_\gamma) - \frac 12 \int_{\R^3}\int_{\R^3} \frac{|\gamma(\br,\br')|^2}{|\br-\br'|} \, d\br \, d\br', 
$$
where, as above, $\gamma(\br,\br')$ denotes the integral kernel of $\gamma$.

\medskip

The existence of a solution to (\ref{eq:minmod}) for a neutral or positively charged system is established in~\cite{Solovej-91} for the Hartree model ($E^{\rm xc}=0$), in \cite{LiebHFg} for the Hartree-Fock model, in~\cite{AnaCan09} for the X$\alpha$ and the standard LDA model, and in \cite{LiebMuller} for the M\"uller DMFT functional. 

\medskip

The key-property allowing for a comprehensive mathematical analysis of the bulk limit for the Hartree model is that the ground state {\em density} is unique (which is not the case for the other models presented in this section). This means that in the Hartree framework, all the minimizers to (\ref{eq:minmod}) share the same density. This follows from the fact that the ground state Hartree density solves the variational problem
\begin{equation} \label{eq:rHFdensity}
\inf \left\{{\cal E}(\rho)  , \; \rho \ge 0, \; \sqrt{\rho} \in H^1(\R^3), \; \int_{\R^3} \rho=N \right\},
\end{equation} 
where
$$
{\cal E}(\rho) = F(\rho) +\int_{\R^3}\rho V_{\rho^{\rm nuc}} + \frac 12 D(\rho,\rho)
$$
and
$$
F(\rho) = \inf \left\{\tr\left(-\frac 12 \Delta \gamma\right), \; \gamma \in {\cal S}(L^2(\R^3), \; 0 \le \gamma \le 1,
  \; \tr(\gamma) = N, \; \tr(-\Delta\gamma) < \infty, \; \rho_\gamma=\rho \right\}.
$$
As the functional ${\cal E}(\rho)$ is strictly convex on the convex set
$$
\left\{\rho \ge 0, \; \sqrt{\rho} \in H^1(\R^3), \; \int_{\R^3} \rho=N \right\},
$$
uniqueness follows.

\medskip

The Euler equation for the Hartree model reads
\begin{equation}\label{eq:EulerHartree}
\left\{ \begin{array}{l}
\dps \gamma^0 = \sum_{i=1}^{+\infty} n_i |\phi_i\rangle \langle \phi_i|, \quad \rho^0(\br) = \rho_{\gamma^0}(\br) = \sum_{i=1}^{+\infty} n_i |\phi_i(\br)|^2, \\
H^0 \phi_i = \epsilon_i \phi_i, \quad \dps \langle \phi_i|\phi_j\rangle = \delta_{ij}, \\ \dps 
n_i = 1 \mbox{ if } \epsilon_i < \epsilon_{\rm F}, \;  0 \le n_i \le 1 \mbox{ if } \epsilon_i = \epsilon_{\rm F}, \;  n_i=0  \mbox{ if } \epsilon_i > \epsilon_{\rm F}, \quad \sum_{i=1}^{+\infty} n_i= N, \\
\dps H^0 = -\frac 12 \Delta +V^0, \\
-\Delta V^0 = 4\pi(\rho^{\rm nuc}-\rho^0). 
\end{array} \right.
\end{equation}
It can be proved that the essential spectrum of the self-adjoint operator $H^0$ is equal to $\R_+$ and that, for a neutral or positively charged system, $H^0$ has at least $N$ negative eigenvalues. The scalar $\epsilon_{\rm F}$, called the Fermi level, can be interpreted as the Lagrange multiplier of the constraint $\tr(\gamma^0)=N$.

Assuming that $\epsilon_N < \epsilon_{N+1}$, the ground state density operator $\gamma^0$ of the Hartree model is unique: It is the orthogonal projector
$$
\gamma^0 = \sum_{i=1}^N |\phi_i\rangle \langle\phi_i|.
$$
In this case, (\ref{eq:EulerHartree}) can be rewritten under the more compact form
\begin{equation}\label{eq:EulerHartree2}
\left\{ \begin{array}{l}
\gamma^0 = 1_{(-\infty,\rm \epsilon_{\rm F}]}(H^0), \quad \rho^0=\rho_{\gamma^0}, \\
\dps H^0=-\frac 12 \Delta +V^0, \\
-\Delta V^0 = 4\pi(\rho^{\rm nuc}-\rho^0), 
\end{array} \right.
\end{equation}
for any $\epsilon_N < \epsilon_{\rm F} < \epsilon_{N+1}$. In this equation, the notation $1_{(-\infty,\rm\epsilon_F]}(H^0)$ is used for the spectral projector of $H^0$ corresponding to the spectrum in the interval $(-\infty,\rm\epsilon_F]$.

\medskip

Lastly, we remark that if smeared nuclei are used, then $D(\rho^{\rm nuc}_{\rm per},\rho^{\rm nuc}_{\rm per})$ is well defined (and finite). This allows us to reformulate the Hartree ground state problem as
\begin{equation}\label{eq:minHartree}
\inf \left\{ \widetilde E_{\rho^{\rm nuc}}^{\rm Hartree}(\gamma), \; \gamma \in {\cal S}(L^2(\R^3)), \; 0 \le \gamma \le 1, \; \tr(\gamma) = N, \; \tr(-\Delta \gamma) < \infty \right\},
\end{equation}
where
$$
\widetilde E_{\rho^{\rm nuc}}^{\rm Hartree}(\gamma) =  \tr \left( -\frac 12 \Delta \gamma \right) + \frac 12 D(\rho^{\rm nuc}-\rho_\gamma,\rho^{\rm nuc}-\rho_\gamma).
$$
The main interest of this new formulation of the Hartree problem is that the functional $\widetilde E_{\rho^{\rm nuc}}^{\rm Hartree}$ is the sum of two non-negative contributions: the kinetic energy and the Coulomb energy of the total charge distribution $\rho^{\rm nuc}-\rho_\gamma$. The presence of the unphysical terms corresponding to the self-interaction of nuclei in $D(\rho^{\rm nuc}_{\rm per},\rho^{\rm nuc}_{\rm per})$ is not a problem for our purpose.

\medskip

The time-dependent version of the Hartree model formally reads
$$
i \frac{d\gamma}{dt}(t) = \left[ -\frac 12 \Delta - (\rho^{\rm nuc}(t)-\rho_{\gamma(t)}) \star |\cdot|^{-1},\gamma(t) \right],
$$
where $[A,B]=AB-BA$ denotes the commutator of the operators $A$ and $B$. We are not going to elaborate further on the precise mathematical meaning of this formal equation for finite systems, but refer the reader to~\cite{Arnold} and references therein (see in particular~\cite[Section~XVII.B.5]{DautrayLions}) for further precision on the mathematical meaning of the above equation. On the other hand, we will define and study a mild version of it in the case of crystals with defects in Section~\ref{sec:TDexpansion}.

\section{The Hartree model for crystals}
\label{sec:crystals}

The Hartree model presented in the previous section 
describes a \emph{finite} system of $N$ electrons in
the electrostatic potential created by a nuclear density of charge $\rho^{\rm
  nuc}$. Our goal is to describe an \emph{infinite} crystalline material
obtained in the bulk limit. In fact we shall
consider two such systems. The first one is the periodic crystal obtained when, in the bulk limit, the nuclear density approaches the periodic nuclear distribution of the perfect
crystal:
\begin{equation}
 \rho^{\rm nuc}\rightarrow \rho_{\rm per}^{\rm nuc},
\label{case_periodic}
\end{equation}
$\rho^{\rm nuc}_{\rm per}$ being a $\cR$-periodic distribution. The set $\cR$
is a periodic lattice of $\R^3$:
\begin{equation}
  \label{eq:lattice_R}
  \cR = \Z {\bold a}_1 + \Z {\bold a}_2 + \Z {\bold a}_3,
\end{equation}
where $({\bold a}_1,{\bold a}_2,{\bold a}_3)$ is a given triplet of linearly independent vectors of $\R^3$. The second system
is the previous crystal in the presence of a local defect: 
\begin{equation}
\rho^{\rm nuc}\rightarrow \rho_{\rm per}^{\rm nuc}+\mnu,
\label{case_defect}
\end{equation}
$\mnu$ representing the nuclear charge of the defect. The functional spaces in which $\rho^{\rm nuc}_{\rm per}$ and $\mnu$ are chosen are made precise below.

\subsection{Basics of Fourier and Bloch-Floquet theories}

A perfect crystal is characterized by a lattice $\cR$ of $\R^3$ and a $\cR$-periodic nuclear charge distribution $\rho^{\rm nuc}_{\rm per}$. Not surprisingly, Fourier and Bloch-Floquet theories, which allow to conveniently exploit the periodicity of the problem, play essential roles in the mathematical description of the electronic structure of crystals. 

\medskip

Let $\cR^\ast$ be the reciprocal lattice of the lattice $\cR$ defined in~\eqref{eq:lattice_R} 
(also called dual lattice):
$$
\cR^\ast = \Z {\bold a}_1^\ast + \Z {\bold a}_2^\ast + \Z {\bold a}_3^\ast, \quad \mbox{where} \quad  {\bold a}_i \cdot {\bold a}_j^\ast = 2\pi \delta_{ij}.
$$ 
Denote by $\Gamma$ a unit cell of~$\cR$. Recall that a unit cell is a semi-open bounded polytope of $\R^3$ such that the cells $\Gamma+\bR = \left\{(\br+\bR), \, \br \in \Gamma\right\}$ for $\bR \in \cR$ form a tessellation of the space $\R^3$ (\textit{i.e.} $(\Gamma+\bR) \cap (\Gamma+\bR')=0$ if $\bR \neq \bR'$ and $\cup_{\bR \in \cR} (\Gamma+\bR) =\R^3$). A possible choice for $\Gamma$ is $\left\{x_1{\bold a}_1 + x_2 {\bold a}_2 + x_3 {\bold a}_3, \; -1/2 < x_i \le 1/2 \right\}$. Another choice is the Wigner-Seitz cell of $\cR$, which is by definition the semi-open Voronoi cell of the origin for the lattice $\cR$. Lastly, we denote by $\Gamma^\ast$ the first Brillouin zone, that is the Wigner-Seitz cell of the dual lattice. Let us illustrate these concepts on the simplest example, the cubic lattice, for which $\cR=a\Z^3$ (for some $a > 0$). In this particular case, $\cR^\ast=\frac{2\pi}a\Z^3$, the Wigner-Seitz cell is $\Gamma=(-a/2,a/2]^3$ and $\Gamma^\ast=(-\pi/a,\pi/a]^3$.

For each $\bK \in \cR^\ast$, we denote by $e_\bK(\br) = |\Gamma|^{-1/2} \mathrm{e}^{i\bK\cdot\br}$ the Fourier mode with wavevector $\bK$. According to the theory of Fourier series, each $\cR$-periodic distribution $v$ can be expanded in Fourier series as
\begin{equation} \label{eq:Fourier}
v = \sum_{\bK \in \cR^\ast} c_\bK(v) \, e_\bK,
\end{equation}
where $c_{\bK}(v)$ is the $\bK$-th Fourier coefficient of $v$, the convergence of the series holding in the distributional sense. We introduce the usual $\cR$-periodic $L^p$ spaces defined by
$$
L^p_{\rm per}(\Gamma) := \left\{ \left. v \in L^p_{\rm loc}(\mathbb{R}^3) \; \right|\; v \; \mathcal{R}\mbox{-periodic} \right\},
$$
and endow them with the norms
$$
\|v\|_{L^p_{\rm per}(\Gamma)} := \left( \int_\Gamma |v|^p \right)^{1/p} \quad \mbox{for } 1 \le p < \infty \quad \mbox{and} \quad \|v\|_{L^\infty_{\rm per}(\Gamma)} := \mbox{ess-sup}\,|v|.
$$
In particular,
$$ 
\|v\|_{L^2_{\rm per}(\Gamma)}=(v,v)_{L^2_{\rm per}(\Gamma)}^{1/2} \quad \mbox{where} \quad 
\left( v, w \right)_{L^2_{\rm per}(\Gamma)} := \int_{\Gamma} \overline{v} w.
$$
Any function $v\in L^2_{\rm per}(\Gamma)$ can be expanded in Fourier modes according to (\ref{eq:Fourier}), the Fourier coefficients being given by the simple formula
$$ 
c_\bK(v) = \frac{1}{|\Gamma|^{1/2}}\int_{\Gamma} v(\br) \, \mathrm{e}^{-i\bK\cdot\br}\,d\br,
$$
and the convergence of the series (\ref{eq:Fourier}) also holds in $L^2_{\rm per}(\Gamma)$. Besides,
$$
\forall (v,w) \in L^2_{\rm per}(\Gamma) \times L^2_{\rm per}(\Gamma), \quad \left( v, w \right)_{L^2_{\rm per}(\Gamma)} = \sum_{\bK\in \cR^\ast}  \overline{c_\bK(v)}c_\bK(w).
$$
For each $s \in \R$, the $\cR$-periodic Sobolev space of index $s$ is defined as
$$
H^s_{\rm per}(\Gamma) := \left\{ \left. v = \sum_{\bK \in \cR^\ast} c_\bK(v) e_\bK \; \right |\; \sum_{\bK \in \cR^\ast} (1+|\bK|^2)^s|c_\bK(v)|^2 < \infty \right\},
$$
and endowed with the inner product
$$
(v,w)_{H^s_{\rm per}(\Gamma)} := \sum_{\bK\in\mathcal{R}^*}(1+|\bK|^2)^s \overline{c_\bK(v)}c_\bK(w).
$$

\medskip

The Bloch-Floquet theory was introduced by Floquet for periodic differential equations and generalized by Bloch to periodic partial differential equations. We just recall the basic results of this theory used in this proceeding and refer the reader to~\cite{ReeSim4} for further precisions. 

Any function $f \in L^2(\R^3)$ can be decomposed by the Bloch-Floquet transform as 
$$
f(\br) = \fint_{\Gamma^\ast} f_\bq(\br) \, \mathrm{e}^{i \bq \cdot \br} d\bq,
$$
where $\fint_{\Gamma^\ast}$ is a notation for $|\Gamma^\ast|^{-1}\int_{\Gamma^\ast}$ and where the functions $f_\bq$ are defined by 
\begin{equation}
f_\bq(\br)=\sum_{\bR \in\cR}f(\br+\bR) \mathrm{e}^{-i\bq \cdot (\br+\bR)}=\frac{(2\pi)^{3/2}}{|\Gamma|}
\sum_{\bK\in\cR^\ast}\widehat{f}(\bq+\bK)\mathrm{e}^{i\bK\cdot\br}.
\label{eq:def_Bloch_Floquet}
\end{equation}
For almost all $\bq \in \R^3$, $f_\bq \in L^2_{\rm per}(\Gamma)$. Besides, 
$f_{\bq+\bK}(\br)=f_\bq(\br)\mathrm{e}^{-i\bK\cdot\br}$ for all $\bK \in \cR^\ast$ and almost all $\bq \in \R^3$. Lastly,
$$
\|f\|_{L^2(\R^3)}^2 = \fint_{\Gamma^\ast} \|f_\bq\|_{L^2_{\rm per}(\Gamma)}^2 \, d\bq.
$$
For $\bR \in \R^3$, we denote by $\tau_\bR$ the translation operator defined by 
$$
\forall v \in L^2(\R^3), \quad (\tau_\bR v)(\br) = v(\br-\bR).
$$
The main interest of the Bloch-Floquet transform (\ref{eq:def_Bloch_Floquet}) is that it provides a ``block diagonalization'' of any $\cR$-periodic operator, that is of any operator on $L^2(\R^3)$ which commutes with $\tau_\bR$ for all $\bR \in \cR$. Consider first a bounded $\cR$-periodic operator $A$ on $L^2(\R^3)$. Then there exists a family $(A_\bq)_{\bq \in \Gamma^\ast}$ of bounded operators on $L^2_{\rm per}(\Gamma)$ such that 
\begin{equation} \label{eq:opB}
\forall v \in L^2(\R^3), \quad (Av)_\bq = A_\bq v_\bq \quad \mbox{for almost all } q \in \Gamma^\ast.
\end{equation}
If, in addition, $A$ is self-adjoint on $L^2(\R^3)$, then $A_\bq$ is self-adjoint on $L^2_{\rm per}(\Gamma)$ for almost all $\bq \in \Gamma^\ast$ and 
$$
\sigma(A) = \overline{\bigcup_{\bq \in \Gamma^\ast} \sigma(A_\bq)}.
$$
In particular, the translation operators $(\tau_\bR)_{\bR \in \cR}$, which obviously commute with each other, are homotheties in the Bloch-Floquet representation 
$$
\forall \bR \in \cR, \quad 
(\tau_\bR)_\bq = \mathrm{e}^{i\bq \cdot \bR} 1_{L^2_{\rm per}(\Gamma)}.
$$

As $(e_\bK)_{\bK \in \cR^\ast}$ form an orthonormal basis of $L^2_{\rm per}(\Gamma)$, it follows from (\ref{eq:opB}) that any bounded $\cR$-periodic operator on $L^2(\R^3)$ is completely characterized by the Bloch-Floquet matrices $(([A_{\bK,\bK'}(\bq)])_{(\bK,\bK') \in \cR^\ast\times\cR^\ast})_{\bq \in \Gamma^\ast}$ defined for almost all $\bq \in \Gamma^\ast$ by
$$
A_{\bK,\bK'}(\bq) := \langle e_\bK, A_\bq e_{\bK'} \rangle_{L^2_{\rm per}(\Gamma)}.
$$
In particular, it holds
$$
\forall v \in L^2(\R^3), \quad \widehat{(Av)}(\bq+\bK)= \sum_{\bK'\in \cR^\ast} A_{\bK,\bK'}(\bq) \widehat v(\bq+\bK'),
$$
for all $(\bK,\bK') \in \cR^\ast \times \cR^\ast$ and almost all $\bq \in \Gamma^\ast$.

\begin{figure}[h]
\centering
\includegraphics[width=10cm]{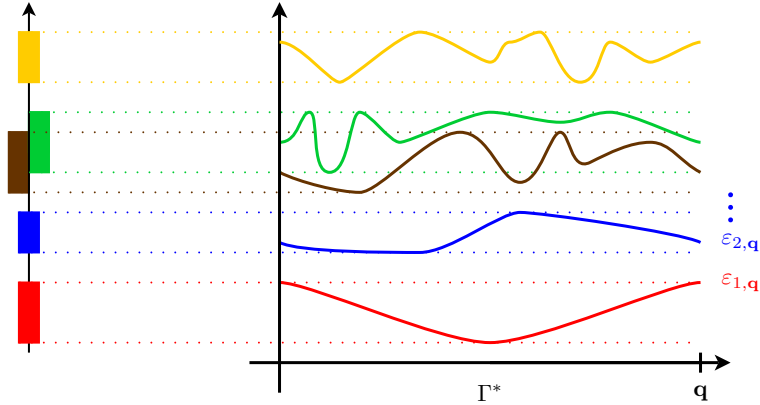}
\caption{\label{fig:Bloch_bands} The spectrum of a periodic Schr\"odinger operator is a union of bands, as a consequence of the Bloch-Floquet decomposition.}
\end{figure}

For unbounded operators, the situation is a little bit more intricate. Let us limit ourselves to the case of$\cR$-periodic Schr\"odinger operators of the form 
$$
H = -\frac 12 \Delta + V_{\rm per}
$$
with $V_{\rm per} \in L^2_{\rm per}(\Gamma)$. By the Kato-Rellich theorem and~\cite[Theorem~XIII.96]{ReeSim4}, the operator $H$ is self-adjoint on $L^2(\R^3)$, with domain $H^2(\R^3)$. It can also be decomposed as follows:
$$
\forall v  \in H^2(\R^3), \quad v_\bq \in H^2_{\rm per}(\Gamma) \quad \mbox{and} \quad (Hv)_\bq = H_\bq v_\bq \quad \mbox{for almost all } \bq \in \Gamma^\ast,
$$
where $H_\bq$ is the self-adjoint operator on $L^2_{\rm per}(\Gamma)$ with domain $H^2_{\rm per}(\Gamma)$, defined by
$$
H_\bq = -\frac{1}{2} \Delta-i\bq \cdot\nabla +\frac{|\bq|^2}{2} +V_{\rm per}.
$$ 
It is easily seen that for each $\bq \in \Gamma^\ast$, $H_\bq$ is bounded below and has a compact resolvent. Consequently, there exists a sequence $(\epsilon_{n,\bq})_{n \ge 1}$ of real numbers going to $+\infty$, and an orthonormal basis $(u_{n,_\bq})_{n \ge 1}$ of $L^2_{\rm per}(\Gamma)$ such that
$$
H_\bq  = \sum_{n=1}^{+\infty} \epsilon_{n,\bq} |u_{n,\bq}\rangle \langle u_{n,\bq}|.
$$
As the mapping $\bq \mapsto H_\bq$ is polynomial on $\R^3$, it is possible to number the eigenvalues $\epsilon_{n,\bq}$ in such a way that $(\epsilon_{n,0})_{n \ge 1}$ is non-decreasing and that for each $n \ge 1$, the mapping $\bq \mapsto \epsilon_{n,\bq}$ is analytic in each direction. Then (see Fig. \ref{fig:Bloch_bands})
\[
\sigma(H) =  \overline{\bigcup_{\bq \in \Gamma^\ast} \sigma(H_\bq)} 
= \bigcup_{n \ge 1} \left[\Sigma_n^-,\Sigma_n^+ \right],
\]
with
\begin{equation}
  \label{eq:band_edges_Sigma}
  \Sigma_n^-=\min_{\bq \in \overline{\Gamma^\ast}}\epsilon_{n,\bq}, \quad \Sigma_n^+=\max_{\bq \in \overline{\Gamma^\ast}}\epsilon_{n,\bq}. 
\end{equation}
The interval $\left[\Sigma_n^-,\Sigma_n^+ \right]$ is called the $n^{\rm th}$ band of the spectrum of $H$. It is possible to prove that the spectrum of $H$ is purely absolutely continuous~\cite{Thomas}. In particular, $H$ has no eigenvalues.

\subsection{Perfect crystals}
\label{sec:perfect}

The purpose of this section is to formally construct, then justify with mathematical arguments, a Hartree model for the electronic structure of perfect crystals. 

As announced, we begin with a formal argument and consider a sequence of finite nuclear distribution $(\rho^{\rm nuc}_n)_{n \in \N}$ converging to the periodic distribution $\rho^{\rm nuc}_{\rm per}$ of the perfect crystal when $n$ goes to infinity. For instance, we can take 
$$
\rho^{\rm nuc}_n = \rho^{\rm nuc}_{\rm per} \left( \sum_{\bR \in \cR \, | \, |\bR| \le n} 1_{\Gamma+\bR} \right)
$$
(we assume that the function describing the nuclear charge in the unit cell of the perfect crystal is supported in some compact set included in the interior of $\Gamma$). We solve the Hartree problem for each $\rho^{\rm nuc}_n$ with the constraint that the system remains neutral for each $n$. Assuming that when $n$ goes to infinity, 
\begin{itemize}
\item the Hartree ground state density converges to some $\cR$-periodic density $\rho^0_{\rm per} \in L^1_{\rm per}(\Gamma)$;
\item the Coulomb potential generated by the total charge converges to some $\cR$-periodic potential $V^0_{\rm per}$;
\item the Hartree ground state density operator converges to some operator $\gamma^0_{\rm per}$;
\item the Fermi level converges to some $\epsilon_{\rm F}^0 \in \R$,
\end{itemize}
we obtain by {\em formally} passing to the limit in (\ref{eq:EulerHartree2}), the self-consistent equations
\begin{equation}\label{eq:EulerHartreePer}
\left\{ \begin{array}{l}
\gamma^0_{\rm per} = 1_{(-\infty,\rm \epsilon^0_{\rm F}]}(H^0_{\rm per}), 
\quad \rho^0_{\rm per}=\rho_{\gamma^0_{\rm per}}, \\
\dps H^0_{\rm per}=-\frac 12 \Delta +V^0_{\rm per}, \\
-\Delta V^0_{\rm per} = 4\pi(\rho^{\rm nuc}_{\rm per}-\rho^0_{\rm per}). 
\end{array} \right.
\end{equation}
Let us comment on this system of equations. First, we notice that for the periodic Coulomb equation $-\Delta V^0_{\rm per} = 4\pi(\rho^{\rm nuc}_{\rm per}-\rho^0_{\rm per})$ to have a solution, each unit cell must be neutral:
\begin{equation} \label{eq:neutral}
\int_{\Gamma} \rho^0_{\rm per} = \int_{\Gamma} \rho^{\rm nuc}_{\rm per} = Z,
\end{equation}
where $Z$ is the number of electrons, and also the number of protons, per unit cell. Second, as $V^0_{\rm per}$ is $\cR$-periodic (and belongs to $L^2_{\rm per}(\Gamma)$ even for point-like nuclei), we can apply the result of the previous section and write down the Bloch-Floquet decomposition of $H^0_{\rm per}$:
\begin{equation} \label{eq:BlochH0per}
(H^0_{\rm per})_\bq = 
-\frac{1}{2} \Delta-i\bq\cdot\nabla +\frac{|\bq|^2}{2} +V_{\rm per}^0
= \sum_{n=1}^{+\infty} \epsilon_{n,\bq} |u_{n,\bq}\rangle \langle u_{n,\bq}|.
\end{equation}
The operator $\gamma^0_{\rm per}=1_{(-\infty,\rm \epsilon^0_{\rm F}]}(H^0_{\rm per})$ then is a bounded self-adjoint operator which commutes with the translations $(\tau_\bR)_{\bR\in \cR}$, and its Bloch-Floquet decomposition reads
$$
(\gamma^0_{\rm per})_\bq = \sum_{n =1}^{+\infty} 1_{\epsilon_{n,\bq}\le \epsilon_{\rm F}^0}  |u_{n,\bq}\rangle \langle u_{n,\bq}|.
$$
Actually, the set $\left\{q \in \Gamma^\ast \, | \, \exists n \ge 1 \mbox{ s.t. } \epsilon_{n,\bq} = \epsilon^0_{\rm F} \right\}$ is of measure zero (the spectrum of $H^0_{\rm per}$ is purely continuous). It follows that $\gamma^0_{\rm per}$ is always an orthogonal projector, even if $\epsilon^0_{\rm F}$ belongs to the spectrum of $H^0_{\rm per}$. 

Using the Bloch decomposition of $\gamma^0_{\rm per}$, we can write the density $\rho^0_{\rm per}$ as
$$
\rho^0_{\rm per}(\br) = \fint_{\Gamma^\ast}  \sum_{n=1}^{+\infty} 1_{\epsilon_{n,q}\le \epsilon_{\rm F}^0}  |u_{n,\bq}(\br)|^2 \, d\bq.
$$
Integrating on $\Gamma$, and using (\ref{eq:neutral}) and the orthonormality of the functions $(u_{n,\bq})_{n\ge 1}$ in $L^2_{\rm per}(\Gamma)$, we obtain
\begin{equation}
Z= \frac{1}{|\Gamma^*|} \sum_{n=1}^{+\infty} \left|\{\bq\in\Gamma^\ast\ |\
  \epsilon_{n,\bq}\leq{\epsilon^0_{\rm F}}\}\right|.
\label{def_Z}
\end{equation}
This equation determines the value of the Fermi level $\rm \epsilon_F$ uniquely.
It is easy to see that if the periodic Coulomb potential is shifted by a uniform constant $C$, and if $\epsilon^0_{\rm F}$ is replaced with $\epsilon^0_{\rm F}+C$, then $\gamma^0_{\rm per}$ and $\rho^0_{\rm per}$ remain unchanged. 

\medskip

The formal bulk limit argument presented above has been rigorously founded by Catto, Le Bris and Lions in \cite{CatBriLio-01}, for $\rho^{\rm nuc}_{\rm per} = \sum_{\bR \in \Z^3} \chi(\cdot-\bR)$ (smeared nuclei of unit charge disposed on the cubic lattice $\Z^3$). It is also possible to justify the periodic Hartree model by passing to the limit on the supercell model with artificial periodic boundary conditions (see \cite{CanDelLew-08a}). The latter approach is less physical, but technically much easier, and its extension to arbitrary crystalline structures (including point-like nuclei) is straightforward. It results from these mathematical works that the Hartree model for perfect crystals is well-defined. More precisely: 
\begin{enumerate}
\item The Hartree ground state density operator $\gamma^0_{\rm per}$ and density $\rho^0_{\rm per}$ of a crystal with periodic nuclear density $\rho^{\rm nuc}_{\rm per}$ (composed of point-like or smeared nuclei) are uniquely defined;
\item The ground state density $\rho^0_{\rm per}$ satisfies the neutrality charge constraint (\ref{eq:neutral});
\item The periodic Coulomb potential $V^0_{\rm per}$ and the Fermi level $\epsilon^0_{\rm F}$ are uniquely defined up to an additive constant (and $V^0_{\rm per}-\epsilon^0_{\rm F}$ is uniquely defined);
\item The ground state density operator $\gamma^0_{\rm per}$ is an infinite rank orthogonal projector satisfying the self-consistent equation (\ref{eq:EulerHartreePer});
\item $\gamma^0_{\rm per}$ can be obtained by minimizing some periodic model set on the unit cell $\Gamma$ (see \cite{CatBriLio-01} for details).
\end{enumerate}

\medskip

In the remainder of the paper we assume that the system is an insulator
(or a semi-conductor) in the sense that the $N^{\rm th}$ band is
strictly below the $(N+1)^{\rm st}$ band:
$$
\Sigma_N^+ < \Sigma_{N+1}^-, 
$$
where $\Sigma_n^\pm$ are defined in~\eqref{eq:band_edges_Sigma}.
In this case, one can choose for $\epsilon_{\rm F}^0$ any number in the range
$(\Sigma_N^+,\Sigma_{N+1}^-)$. The electronic state of the perfect crystal is the same whatever the value of $\epsilon_{\rm F}^0$ in the gap $(\Sigma_N^+,\Sigma_{N+1}^-)$. On the other hand, as will be seen in the next section, fixing the value of $\epsilon_{\rm F}^0$ may change the electronic state of the crystal in the presence of a local defect.

In this paper however, we are only interested in the dielectric response of the crystal, which corresponds to the limit of small defects (in a sense that will be made precise later), and in this limit, the value of $\epsilon_{\rm F}^0$ does not play any role as long as it remains inside the gap $(\Sigma_N^+,\Sigma_{N+1}^-)$. For simplicity, we consider in the following 
$$
\epsilon_{\rm F}^0=\frac{\Sigma_N^++\Sigma_{N+1}^-}{2}.
$$
Lastly, we denote by 
\begin{equation}
  \label{eq:band_gap}
  g = \Sigma_{N+1}^- - \Sigma_N^+ > 0
\end{equation}
the band gap.

\subsection{Crystals with local defects}
\label{sec:RHFperturbed}

We now describe the results of \cite{CanDelLew-08a} dealing with the modelling of local defects in crystals in the framework of the Hartree model. The main idea is to seek the ground state density operator of a crystal with a local defect characterized by the nuclear charge distribution (\ref{case_defect}) under the form
$$
\gamma_{\mnu,\epsilon_{\rm F}^0} = \gamma^0_{\rm per}+Q_{\mnu,\epsilon_{\rm F}^0}.
$$
In this formalism, the defect is seen as a quasi-molecule with nuclear charge distribution $\mnu$ and electronic ground state density operator $Q_{\mnu,\epsilon_{\rm F}^0}$ (and ground state electronic density $\rho_{Q_{\mnu,\epsilon_{\rm F}^0}}$), embedded in the perfect crystal. Here, the charge of the defect is controlled by the Fermi level (the chemical potential). The dual approach, in which the charge of the defect is imposed, is also dealt with in \cite{CanDelLew-08a}. It should be noticed that neither $\mnu$ nor $\rho_{Q_{\mnu,\epsilon_{\rm F}^0}}$ are {\it a priori} non-negative. For instance, the nuclear distribution of a defect corresponding to the replacement of a nuclear of charge $z$ located at point $\bR \in \R^3$ with a nucleus of charge $z'$ is $\mnu = (z'-z)\delta_{\bR}$ and can therefore be positively or negatively charged depending on the value of $z'-z$. Regarding the electronic state, the constraints $(\gamma_{\mnu,\epsilon_{\rm F}^0})^\ast=\gamma_{\mnu,\epsilon_{\rm F}^0}$, $0 \le \gamma_{\mnu,\epsilon_{\rm F}^0} \le 1$ and $\rho_{\gamma_{\mnu,\epsilon_{\rm F}^0}} \ge 0$, respectively read $(Q_{\mnu,\epsilon_{\rm F}^0})^\ast=Q_{\mnu,\epsilon_{\rm F}^0}$, $- \gamma^0_{\rm per} \le Q_{\mnu,\epsilon_{\rm F}^0} \le 1-\gamma^0_{\rm per}$
and $\rho_{Q_{\mnu,\epsilon_{\rm F}^0}} \ge -\rho^0_{\rm per}$.

\medskip

The next step is to exhibit a variational model allowing to compute $Q_{\mnu,\epsilon_{\rm F}^0}$ from $\mnu$, $\epsilon_{\rm F}^0$ and the ground state of the perfect crystal.

First, we perform the following formal calculation of the difference between the Hartree free energy of some trial density operator $\gamma=\Gper+Q$ subjected to the nuclear potential generated by $\rho^{\rm nuc}_{\rm per}+\mnu$, and the Hartree free energy of the perfect crystal:
\begin{eqnarray}
&& \left(\widetilde E^{\rm Hartree}_{\rho^{\rm nuc}_{\rm per}+\mnu}(\Gper+Q)-\epsilon^0_{\rm F}\tr(\Gper+Q)\right) - \left(\widetilde E^{\rm Hartree}_{\rho^{\rm nuc}_{\rm per}}(\Gper)-\epsilon^0_{\rm F}\tr(\Gper)\right) \nonumber \\
&& \dps \qquad \mathop{=}^{\rm formal} \quad \tr\left(-\frac 12 \Delta Q\right) + \int_{\R^3} \rho_Q V^0_{\rm per} - \int_{\R^3}  \rho_Q V_\mnu + \frac 12 D(\rho_Q,\rho_Q)-\epsilon^0_{\rm F}\tr(Q) \nonumber \\
&&  \qquad \qquad \qquad -  \int_{\R^3}  \mnu V^0_{\rm per} + \frac 12 D(\mnu,\mnu).
\label{eq:formalExp}
\end{eqnarray}
The last two terms are constants that we can discard.
Of course, the left-hand side of (\ref{eq:formalExp}) does not have any mathematical sense since it is the difference of two energies both equal to plus infinity. On the other hand, we are going to see that it is possible to give a mathematical meaning to the sum of the first five terms of the right-hand side when $Q$ belongs to some functional space ${\cal Q}$ defined below, and to characterize the ground state density operator $Q_{\mnu,\epsilon_{\rm F}^0}$ of the quasi-molecule, by minimizing the so-defined energy functional on a closed convex subset ${\cal K}$ of ${\cal Q}$.

For this purpose, we first need to extend the definition (\ref{eq:defCoulomb}) of the Coulomb interaction to the Coulomb space $\cC$ defined as  
$$
\cC:=\left\{ f\in \cS'(\R^3)\ \left|\ \widehat f \in L^1_{\rm loc}(\R^3), \, D(f,f):= 4\pi \int_{\R^3} \frac{|\hat f(k)|^2}{|k|^2} \, dk \right. \right\},
$$
where $\cS'(\R^3)$ is the space of tempered distributions on~$\R^3$.
Endowed with its natural inner product
\begin{equation} \label{eq:defCoulombExtension}
\langle f,g \rangle_\cC := D(f,g):= 4\pi \int_{\R^3}
\frac{\overline{\hat f(k)} \, \hat g(k)}{|k|^2} \, dk,
\end{equation}
$\cC$ is a Hilbert space. It can be proved that $L^{6/5}(\R^3) \hookrightarrow \cC$ and that for any $(f,g) \in  L^{6/5}(\R^3) \times L^{6/5}(\R^3)$, it holds
$$
4\pi \int_{\R^3}
\frac{\overline{\hat f(k)} \, \hat g(k)}{|k|^2} \, dk = \int_{\R^3} \int_{\R^3} \frac{f(\br)\, g(\br')}{|\br-\br'|} \, d\br \, d\br'.
$$
Hence, the definition (\ref{eq:defCoulombExtension}) of $D(\cdot,\cdot)$ on $\cC$ is consistent with the usual definition~(\ref{eq:defCoulomb}) of the Coulomb interaction when the latter makes sense. The Coulomb space~$\cC$ therefore is the set of charge distributions of finite Coulomb energy. 

Second, we introduce, for an operator $A$ on $L^2(\R^3)$, the notation 
$$
\begin{array}{ll}
A^{--} := \gamma^0_{\rm per} A \gamma^0_{\rm per}, & \qquad 
A^{-+} := \gamma^0_{\rm per} A (1-\gamma^0_{\rm per}), \\[5pt]
A^{+-} := (1-\gamma^0_{\rm per}) A \gamma^0_{\rm per}, & \qquad 
A^{++} := (1-\gamma^0_{\rm per}) A (1-\gamma^0_{\rm per}),
\end{array}
$$
and note that the constraints $Q=Q^\ast$ and  $-\Gper \le Q \le 1-\Gper$ are equivalent to 
\begin{equation} \label{eq:constraintsQ}
Q^\ast = Q, \qquad Q^2 \le Q^{++}-Q^{--}.
\end{equation}
From the second inequality we deduce that it then holds $Q^{--} \le 0$ and $Q^{++} \ge 0$.
Using the fact that $\tr(V^0_{\rm per}Q) = \int_{\R^3} \rho_Q V^0_{\rm per}$, we formally obtain
\begin{eqnarray*}
&& \!\!\!\!\!\!\!\!\!\!\!\!
\tr\left(-\frac 12 \Delta Q\right) + \int_{\R^3} \rho_Q V^0_{\rm per}-\epsilon^0_{\rm F}\tr(Q) = \tr((H^0_{\rm per}-\epsilon^0_{\rm F})Q) \\ && = \tr((H^0_{\rm per}-\epsilon^0_{\rm F})^{++}Q^{++}) + \tr((H^0_{\rm per}-\epsilon^0_{\rm F})^{--}Q^{--}). 
\end{eqnarray*}
We now remark that, by definition of $\gamma^0_{\rm per}$, $(H^0_{\rm per}-\epsilon^0_{\rm F})^{++} \ge 0$ and $(H^0_{\rm per}-\epsilon^0_{\rm F})^{--} \le 0$, so that the right-hand term of the above expression can be rewritten as 
\begin{equation} \label{eq:TT}
\tr( |H^0_{\rm per}-\epsilon^0_{\rm F}|^{1/2}(Q^{++}-Q^{--})|H^0_{\rm per}-\epsilon^0_{\rm F}|^{1/2}).
\end{equation}
The above expression is well defined in $\R_+\cup \left\{+\infty\right\}$ for all $Q$ satisfying the constraints (\ref{eq:constraintsQ}). It takes a finite value if $Q$ is chosen in the vector space 
\begin{eqnarray}
\cQ &=& \big\{ Q \in \gS_2 \; |  \; Q^\ast = Q, \; \; Q^{--} \in \gS_1, \; Q^{++} \in \gS_1, \label{eq:defQ}
 \\ && \qquad
\qquad \quad |\nabla|Q \in \gS_2, \; |\nabla|Q^{--}|\nabla| \in \gS_1,
\; |\nabla|Q^{++}|\nabla| \in \gS_1  \big\}, \nonumber 
\end{eqnarray}
where $\gS_1$ and $\gS_2$ respectively denote the spaces of trace-class and Hilbert-Schmidt operators on $L^2(\R^3)$ (see Appendix for details). Endowed with its natural norm, or with any equivalent norm such as
$$
\|Q\|_\cQ = \|(1+|\nabla|)Q\|_{\gS_2}+\|(1+|\nabla|)Q^{++}(1+|\nabla|)\|_{\gS_1}+\|(1+|\nabla|)Q^{--}(1+|\nabla|)\|_{\gS_1},
$$
$\cQ$ is a Banach space.

\medskip

Before proceeding further, let us comment on the definition of ${\cal Q}$. As the trial density operators $Q$ must satisfy the constraints (\ref{eq:constraintsQ}), it is natural to impose $Q^\ast = Q$. Since $|H^0_{\rm per}-\epsilon^0_{\rm F}|^{1/2}(1+|\nabla|)^{-1}$ is a bounded operator with bounded inverse (see \cite{CanDelLew-08a}), the four conditions $Q^{--} \in \gS_1$, $Q^{++} \in \gS_1$, $|\nabla|Q^{--}|\nabla| \in \gS_1$ and $|\nabla|Q^{++}|\nabla| \in \gS_1$ are necessary and sufficient conditions for  the expression (\ref{eq:TT}) with $Q$ satisfying (\ref{eq:constraintsQ}) being finite. The other constraints imposed to the elements of $\cQ$ (that is, $Q \in \gS_2$ and $|\nabla|Q \in \gS_2$) follow from the fact that for any $Q$ satisfying (\ref{eq:constraintsQ})
\begin{eqnarray*}
\left( Q^{--} \in \gS_1, \; Q^{++} \in \gS_1 \right)  & \quad \Rightarrow \quad &\left( Q^2 \in \gS_1\right) \\
\left(  |\nabla|Q^{--}|\nabla| \in \gS_1,
\; |\nabla|Q^{++}|\nabla| \in \gS_1 \right)  & \quad \Rightarrow \quad &\left( |\nabla|Q^2|\nabla| \in \gS_1 \right).
\end{eqnarray*}
In order to simplify the notation, we set for $Q \in \cQ$,
\begin{align*}
&\tr_0(Q):=\tr(Q^{++}+Q^{--}), \\
&\tr_0((H^0_{\rm per}-\epsilon^0_{\rm F})Q):=\tr( |H^0_{\rm per}-\epsilon^0_{\rm F}|^{1/2}(Q^{++}-Q^{--})|H^0_{\rm per}-\epsilon^0_{\rm F}|^{1/2}).
\end{align*}
An important result is that the linear application $Q \mapsto \rho_Q$ originally defined on the dense subset $\cQ \cap \gS_1$ of $\cQ$ can be extended in a unique way to a continuous linear application 
\begin{eqnarray*}
\cQ & \rightarrow & L^2(\R^3) \cap \cC \\
Q & \mapsto & \rho_Q.
\end{eqnarray*}
Note that the density associated with a generic element of $\cQ$ is not necessarily an integrable function. On the other hand, its Coulomb energy is always finite. 

\medskip

Let $\mnu$ be such that $V_\mnu = (\mnu \star |\cdot|^{-1}) \in \cC'$. Here and in the sequel 
$$
\cC' := \left\{ V \in L^6(\R^3) \, \left| \, \nabla V \in (L^2(\R^3))^3 \right. \right\}
$$
denotes the dual space of $\cC$, endowed with the inner product
$$
\langle V_1,V_2 \rangle_{\cC'} := \frac{1}{4\pi}
\int_{\R^3} \nabla V_1 \cdot \nabla V_2 = 
\frac{1}{4\pi}\int_{\R^3} |k|^2\overline{\hat V_1(k)} \, \hat V_2(k)\, dk.
$$
It follows from the above arguments that the energy functional
$$
E^{\mnu,\epsilon^0_{\rm F}}(Q) = \tr_0((H^0_{\rm per}-\epsilon^0_{\rm F})Q)  - \int_{\R^3}  \rho_Q V_\mnu + \frac 12 D(\rho_Q,\rho_Q)
$$
is well defined on $\cQ$ and that a good candidate for a variational model allowing to compute the ground state density operator $Q_{\mnu,\epsilon_{\rm F}^0}$ is
\begin{equation} \label{eq:model_defaut}
\inf \left\{ E^{\mnu,\epsilon^0_{\rm F}}(Q), \; Q \in \cK\right\}
\end{equation}
where
\begin{equation} \label{eq:defK}
\cK = \big\{ Q \in \cQ \; | \; -\gamma^0_{\rm per} \le Q \le 1 - 
\gamma^0_{\rm per} \big\}.
\end{equation}
Note that $\cK$ is a closed convex subset of $\cQ$.

\medskip

The above formal construction of the model (\ref{eq:model_defaut}) is justified in \cite{CanDelLew-08a} by means of rigorous bulk limit arguments. To summarize the situation, the Hartree ground state density operator of the crystal with nuclear charge density
$\rho^{\rm nuc}_{\rm per} + \mnu$ (the charge of the defect being controlled by the Fermi level) is given by
$$
\gamma = \gamma^0_{\rm per} + Q_{\mnu,\epsilon_{\rm F}^0}
$$
where $Q_{\mnu,\epsilon_{\rm F}^0}$ is obtained by solving (\ref{eq:model_defaut}).

The existence of a Hartree ground state density operator for a crystal with a local defect, as well as the uniqueness of the corresponding density and some other important properties, are granted by the following theorem which gathers several results from \cite{CanDelLew-08a} and \cite{CanLew10}. 

\begin{theorem}
  \label{thm:defaut} 
Let $\mnu$ such that $(\mnu \star |\cdot|^{-1}) \in L^2(\R^3) + \cC'$. Then,
\begin{enumerate}
\item (\ref{eq:model_defaut}) has at least one minimizer $Q_{\mnu,\epsilon_{\rm F}^0}$, and all the minimizers of (\ref{eq:model_defaut}) share the same density $\rho_{\mnu,\epsilon_{\rm F}^0}$;
\item   $Q_{\mnu,\epsilon_{\rm F}^0}$ is solution to the
self-consistent equation
\begin{equation} \label{eq:SCF}
Q_{\mnu,\epsilon_{\rm F}^0} = 
1_{(-\infty,\epsilon_{\rm F}^0)} \left(H^0_{\rm per} + (\rho_{\mnu,\epsilon_{\rm
    F}^0}-\mnu) \star |\cdot|^{-1} \right) -  
1_{(-\infty,\epsilon_{\rm F}^0]} \left(H^0_{\rm per}\right) + \delta,
\end{equation}
where $\delta$ is a finite-rank self-adjoint operator on $L^2(\R^3)$
such that $0 \le \delta \le 1$ and $\mbox{Ran}(\delta) \subset
\mbox{Ker}\left( H^0_{\rm per} + (\rho_{\mnu,\epsilon_{\rm
    F}^0}-\mnu) \star |\cdot|^{-1} - \epsilon^0_{\rm F} \right)$.
\end{enumerate}
\end{theorem}

The interpretation of the Euler equation (\ref{eq:SCF}), which also reads 
$$
\Gper + Q_{\mnu,\EF} = 1_{(-\infty,\EF]}(H^0_{\mnu,\EF})+\delta
$$
with  
$$
H^0_{\mnu,\EF}=\Hper + (\rho_{\mnu,\EF}-\mnu)\star|\cdot|^{-1}, \quad 0 \le \delta \le 1, \quad \mbox{Ran}(\delta) \subset \mbox{Ker}(H^0_{\mnu,\EF}-\EF),
$$
is the following. The mean-field Hamiltonian $H^0_{\mnu,\EF}$ is uniquely defined, since all the minimizers of (\ref{eq:model_defaut}) share the same density $\rho_{\mnu,\epsilon_{\rm F}^0}$. Besides, the operator $(\rho_{\mnu,\EF}-\mnu)\star|\cdot|^{-1}$ being a relatively compact perturbation of $\Hper$, it results from the Weyl theorem (see~\cite[Section~XIII.4]{ReeSim4}) that the Hamiltonians $\Hper$ and $H^0_{\mnu,\EF}$ have the same essential spectra. On the other hand, while $\Hper$ has no eigenvalues,  $H^0_{\mnu,\EF}$ may have a countable number of isolated eigenvalues of finite multiplicities in the gaps as well as below the bottom of the essential spectrum. The only possible accumulation points of these eigenvalues are the edges of the bands.

\begin{figure}[h]
\centering
\includegraphics[width=10cm]{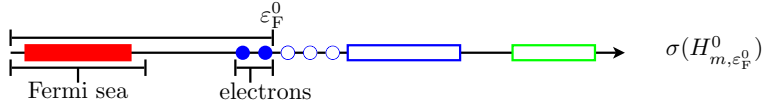}
\caption{\label{fig:spectrum} General form of the spectrum of the self-consistent operator $H^0_{\mnu,\EF}$, in the presence of a defect and for a fixed chemical potential $\epsilon^0_{\rm F}$.}
\end{figure}

If $\EF \notin \sigma(H^0_{\mnu,\EF})$, then $\delta = 0$ and the ground state density operator of the crystal in the presence of the defect is the orthogonal projector $\Gper + Q_{\mnu,\EF}$: All the energy levels lower that the Fermi level are fully occupied while the other ones are empty (see Fig. \ref{fig:spectrum}). In this case, $Q_{\mnu,\EF}$ is both a Hilbert-Schmidt operator and the difference of two projectors. It therefore follows from \cite[Lemma 2]{HaiLewSer-05a} that 
\begin{equation}
\tr_0(Q_{\mnu,\epsilon_{\rm F}^0}) \in \N.
\end{equation}
Assuming that $\mnu \in L^1(\R^3)$ and $\int_{\R^3} \mnu \in \N$, the integer  
$$
\int_{\R^3} \mnu - \tr_0(Q_{\mnu,\epsilon_{\rm F}^0})
$$
can be interpreted as the {\em bare} charge of the defect (in contrast with the \emph{screened} or \emph{renormalized} charge to be defined later).

If $\EF \in \sigma(H^0_{\mnu,\EF})$, then the energy levels with energy $\EF$ may be fully or partially occupied, and it may {\it a priori} happen that  (\ref{eq:model_defaut}) has several minimizers, differing from one another by a finite rank self-adjoint operator with range in $\mbox{Ker}(H^0_{\mnu,\EF}-\EF)$.

\section{Dielectric response of a crystal}
\label{sec:dielectric}

In this section, we study the response of the electronic ground state of
a crystal to a {\em small}, {\em effective} potential. In Section~\ref{sec:expansion}, we consider a time-independent perturbation $V \in L^2(\R^3) + \cC'$, with $\|V\|_{L^2+\cC'} < \alpha$ (for some $\alpha > 0$ small enough). It can be proved (see~\cite[Lemma~5]{CanLew10}) that there exists $\beta > 0$ such that 
\begin{equation} \label{eq:boundab}
\left( \|\mnu \star |\cdot|^{-1}\|_{L^2+\cC'} < \beta \right) \quad \Rightarrow \quad 
\left( \|(\rho_{\mnu,\EF}-\mnu) \star |\cdot|^{-1}\|_{L^2+\cC'} < \alpha \right).
\end{equation}
The results of Section~\ref{sec:expansion} therefore directly apply to the case of a crystal with a local defect with nuclear charge distribution $\mnu$, provided the defect is small enough (in the sense that  $\|\mnu \star |\cdot|^{-1}\|_{L^2+\cC'} < \beta$).

In Section~\ref{sec:TDexpansion}, we consider a time-dependent perturbation 
\begin{equation} 
  \label{eq:TD_pot}
v(t,\br) = (\rho(t,\cdot) \star |\cdot|^{-1})(\br) \qquad  \mbox{with}  
\qquad \rho \in L^1_{\rm loc}(\R,L^2(\R^3)\cap \cC).
\end{equation}

\subsection{Series expansion of the time-independent response}
\label{sec:expansion}

\begin{figure}[h] 
\centering
\includegraphics{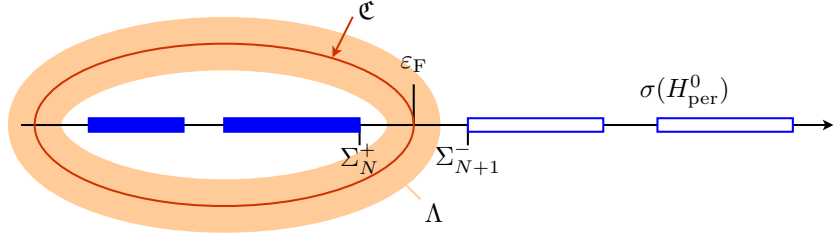}
\caption{Graphical representation of a contour ${\mathfrak C} \subset \C$
  enclosing $\sigma(H^0_{\rm per}) \cap (-\infty,\epsilon_{\rm F}^0]$ and
  of the compact set $\Lambda$.}
\label{fig:contour}
\end{figure}

For $V \in L^2(\R^3) + \cC'$, the spectrum of $H^0_{\rm per}+V$ depends continuously of $V$. In particular (see~\cite[Lemma~2]{CanLew10}), there exists some $\alpha > 0$, such that if $\mathfrak C$ is a smooth curve in the complex plane enclosing the whole spectrum of $H^0_{\rm per}$ below $\epsilon_{\rm F}^0$, crossing the real line at $\epsilon_{\rm F}^0$ and at some $c < \inf\sigma(H^0_{\rm per})$ and such that
$$
d(\sigma(H^0_{\rm per}),\Lambda) = \frac g 4 \quad \mbox{where} \quad 
\Lambda = \left\{z\in \C \; \left| \; d(z,{\mathfrak C}) \le \frac g4 \right. \right\},
$$
$d$ denoting the Euclidean distance in the complex plane and $g$ the
band gap~\eqref{eq:band_gap} (see Fig.~\ref{fig:contour}), then $\sigma(\Hper+V) \cap (-\infty,\EF]$ is contained in the interior of ${\mathfrak C}$ for all $V \in L^2(\R^3)+\cC'$ such that $\|V\|_{L^2+\cC'} < \alpha$.

As a consequence, we obtain that for all $V \in L^2(\R^3)+\cC'$ such that $\|V\|_{L^2+\cC'}~<~\alpha$,
\begin{eqnarray}
Q_V &=& 1_{(-\infty,\epsilon_{\rm F}^0)} \left(H^0_{\rm per} + V \right) -  
1_{(-\infty,\epsilon_{\rm F}^0]} \left(H^0_{\rm per}\right) \nonumber \\
&=& \frac{1}{2i\pi} \oint_{\mathfrak C} \left(\left( z-H^0_{\rm per}-V \right)^{-1} - \left(z-H^0_{\rm
      per}\right)^{-1}\right) \, dz, \label{eq:resolvent}
\end{eqnarray}
where we have used the fact that $\EF \notin \sigma(H^0_{\rm per} + V)$ to establish the first equality, and the Cauchy formula to derive the second one.

Expanding (\ref{eq:resolvent}) in powers of $V$, we obtain  
\begin{equation} \label{eq:expansion}
Q_V = \sum_{n=1}^{N} Q_{n,V}+\widetilde Q_{N+1,V},
\end{equation}
where we have gathered the terms involving powers of $V$ larger than $N$ in a remainder $\widetilde Q_{N+1,V}$. The linear contribution is given by
\begin{equation}
Q_{1,V} =  \frac{1}{2i\pi} \oint_{\mathfrak C}
\left( z-H^0_{\rm per} \right)^{-1} V \left(z-H^0_{\rm
      per}\right)^{-1} \, dz.
\label{eq:Q1V} 
\end{equation}
The higher order contributions and the remainder are respectively given by
$$
Q_{n,V} =  \frac{1}{2i\pi} \oint_{\mathfrak C}
\left( z-H^0_{\rm per} \right)^{-1} \left[ V \left(z-H^0_{\rm
      per}\right)^{-1}\right]^n  \, dz 
$$
and
$$ 
\widetilde Q_{{N+1},V} =  \frac{1}{2i\pi} \oint_{\mathfrak C}
\left( z-H^0_{\rm per}-V \right)^{-1}  \left[ V \left(z-H^0_{\rm
      per}\right)^{-1}\right]^{N+1}   \, dz.
$$

\begin{proposition} 
  \label{prop:chi0} 
  The terms of the perturbation expansion (\ref{eq:expansion}) enjoy the following properties. 
  \begin{enumerate}
  \item The $k$-linear application
    $$
    (V_1,\cdots,V_n) \mapsto  \frac{1}{2i\pi} \oint_{\mathfrak C}
    \left( z-H^0_{\rm per} \right)^{-1}  V_1 \left(z-H^0_{\rm
      per}\right)^{-1} \cdots  V_n \left(z-H^0_{\rm
      per}\right)^{-1}  \, dz 
    $$
    is well-defined and continuous from $(L^2(\R^3)+\cC')^n$ to $\cQ$ for all $n \ge 1$, and from  $(L^2(\R^3)+\cC')^n$ to $\gS_1$ for all $n \ge 6$. In particular, for all $V \in L^2(\R^3) + \cC'$, $Q_{n,V} \in \cQ$ for all $n \ge 1$ and $Q_{n,V} \in \gS_1$ for all $n \ge 6$. Besides, for all  $V \in L^2(\R^3) + \cC'$, $\tr_0(Q_{n,V})=0$ for all $n \ge 1$ and $\tr(Q_{n,V})=0$ for all $n \ge 6$. 
  \item If $V \in L^1(\R^3)$, $Q_{n,V}$ is in $\gS_1$ for each $n \ge 1$ and $\tr(Q_{n,V})=0$.
  \item For each $V\in L^2(\R^3) + \cC'$ such that $\|V\|_{L^2+\cC'}~<~\alpha$, the operator $\widetilde Q_{N+1,V}$ is in $\cQ$ for all $N \ge 0$ with $\tr_0(\widetilde Q_{N+1,V})=0$, and in $\gS_1$ for all $N \ge 5$, with $\tr(\widetilde Q_{N+1,V})=\tr_0(\widetilde Q_{N+1,V})=0$.
  \end{enumerate}
\end{proposition}

We are now in position to define some operators which play an important role in the sequel:
\begin{itemize}
\item the Coulomb operator $v_{\rm c}$, which defines a bijective isometry between $\cC$ and $\cC'$:
  $$
  v_{\rm c}(\rho) := \rho \star |\cdot|^{-1};
  $$
\item the independent particle polarization operator $\chi_0$ defined by
  $$
  \chi_0 (V) := \rho_{Q_{1,V}},
  $$
  which provides the first order response of the electronic density of the crystal to a time-independent modification of the effective potential. The operator $\chi_0$ is a continuous linear application from $L^1(\R^3)$ to $L^1(\R^3)$ and 
  from $L^2(\R^3) + \cC'$ to $L^2(\R^3) \cap \cC$;
\item the linear operator ${\cal L}$ defined by
  $$
  \cL:=-\chi_0v_{\rm c},
  $$
  which is a bounded nonnegative self-adjoint operator on $\cC$. As a consequence, $(1+\cL)^{-1}$ is a well-defined bounded self-adjoint operator on $\cC$;
\item the dielectric operator $\epsilon = v_{\rm c}(1+\cL)v_{\rm c}^{-1}$, and its inverse, the dielectric permittivity operator 
  $$
  \epsilon^{-1} = v_{\rm c}(1+\cL)^{-1}v_{\rm c}^{-1},
  $$
  both being continuous linear operators on $\cC'$. Note that the hermitian dielectric operator, defined as $\widetilde\epsilon= v_{\rm c}^{-1/2}\epsilon v_{\rm c}^{1/2}$ is a self-adjoint, invertible, bounded operator on $L^2(\R^3)$ and is for this reason conveniently used in mathematical proofs.
\end{itemize}

We now focus our attention on the total Coulomb potential 
$$
V_\mnu = (\mnu-\rho_{\mnu,\EF}) \star |\cdot|^{-1} = v_{\rm c}(\mnu-\rho_{\mnu,\EF}),
$$
generated by some charge distribution $\mnu$ such that $\|\mnu\star|\cdot|^{-1}\|_{L^2+\cC'}  < \beta$, and on the response $\rho_{\mnu,\EF}$ of the Fermi sea. In view of (\ref{eq:boundab}), we can apply the above results and deduce from (\ref{eq:expansion})  that
\begin{eqnarray}
\rho_{\mnu,\EF} &=& \rho_{Q_{-V_\mnu}} = \rho_{Q_{1,-V_\mnu}} + \rho_{\widetilde Q_{2,-V_\mnu}} = - \chi_0 V_\mnu + \rho_{\widetilde Q_{2,-V_\mnu}} \nonumber \\
&=& é \cL(\mnu-\rho_{\mnu,\EF})+\rho_{\widetilde Q_{2,-V_\mnu}}.\label{eq:rhonu}
\end{eqnarray}
The above relation, which also reads
\begin{equation}
(\mnu-\rho_{\mnu,\epsilon_{\rm F}^0})=(1+\cL)^{-1}\mnu-(1+\cL)^{-1}(\rho_{\widetilde Q_{2,-V_\mnu}})
\label{eq:SCF_rewritten}
\end{equation}
or 
\begin{equation}
V_\mnu = v_{\rm c}(1+\cL)^{-1}\mnu-v_{\rm c}(1+\cL)^{-1}(\rho_{\widetilde Q_{2,-V_\mnu}}),
\label{eq:SCF_rewritten_2}
\end{equation}
is fundamental since it allows to split the quantities of interest (the total charge $(\mnu-\rho_{\mnu,\epsilon_{\rm F}^0})$ or the total Coulomb potential $V_\mnu$ generated by the defect) into two components:
\begin{itemize}
\item a linear contribution in $\mnu$, very singular, and responsible for charge renormalization at the microscopic level, and for the dielectric properties of the crystal at the macroscopic level;
\item a nonlinear contribution which, in the regime under study ($\|\mnu \star |\cdot|^{-1}\|_{L^2+\cC'} < \beta$), is regular at the microscopic level and vanishes in the macroscopic limit. 
\end{itemize}

\subsection{Properties of $Q_{\mnu,\EF}$ and $\rho_{\mnu,\EF}$ for small amplitude defects}
\label{sec:Qsmall}

The relation (\ref{eq:rhonu}) ,combined with the properties of the operator $\cL$ stated in Proposition~\ref{lem:limL} below, allows to derive some interesting properties of $Q_{\mnu,\EF}$ and $\rho_{\mnu,\EF}$ and to propose a definition of the renormalized charge of the defect. 

\begin{proposition} \label{lem:limL} 
Let $\rho \in L^1(\R^3)$. Then, $\cL(\rho) \in L^2(\R^3) \cap \cC$,
$\widehat{\cL(\rho)}$ is continuous on $\R^3 \setminus
\cR^\ast$, and for all $\sigma \in S^2$ (the unit sphere of $\R^3$), 
\begin{equation} \label{eq:limL}
\lim_{\eta \to 0^+} \widehat{\cL(\rho)}(\eta \sigma) = (\sigma^T L
\sigma) \widehat\rho(0) 
\end{equation}
where $L \in \R^{3 \times 3}$ is the non-negative symmetric matrix
defined by
\begin{equation} \label{eq:mk}
\forall \bk \in \R^3, \quad 
\bk^TL \bk  =   \frac{8\pi}{|\Gamma|}
\sum_{n=1}^N\sum_{n'=N+1}^{+\infty}\fint_{\Gamma^\ast}\frac{\left|\langle(\bk\cdot\nabla_\br)u_{n,\bq},u_{n',\bq}) \rangle_{L^2_{\rm
        per}(\Gamma)}\right|^2}{\big(\epsilon_{n',\bq}-\epsilon_{n,\bq}\big)^3} \, d\bq, 
\end{equation}
where the $\epsilon_{n,\bq}$'s and the $u_{n,\bq}$'s are the eigenvalues and
eigenvectors arising in the spectral decomposition \eqref{eq:BlochH0per} of
$(H^0_{\rm per})_\bq$. Additionally, 
\begin{equation} \label{eq:Lpositive}
L_0 = \frac 13 \tr(L) > 0.
\end{equation}
\end{proposition}

Notice that the convergence of the series (\ref{eq:mk}) is granted by the fact that $\epsilon_{n',\bq}-\epsilon_{n,\bq} \ge \Sigma_{n'}^--\Sigma_n^+ \ge g$ for all $n \le N < n'$ and all $\bq \in \Gamma^\ast$ (where $g > 0$ is the band gap), and the existence of $C \in \R_+$ such that $\|u_{n,\bq}\|_{H^2_{\rm per}(\Gamma)} \le C$ for all $1 \le n \le N$ and all $\bq \in \Gamma^\ast$. Actually, the convergence of the series is rather fast since $\dps\Sigma_{n'}^- \mathop{\sim}_{n'\to \infty} C n'^{2/3}$ (this estimate is obtained by comparing the eigenvalues of $\Hper$ with those of the Laplace operator on $L^2_{\rm per}(\Gamma)$).

We do not reproduce here the quite technical proof of Proposition~\ref{lem:limL}. Let us however emphasize the essential role played by the long range character of the Coulomb potential. If $|\cdot|^{-1}$ is replaced by a potential $v_r\in L^1(\R^3)$, then for all $\rho \in L^1(\R^3)$, $\rho\star v_r\in L^1(\R^3)$, hence $\cL(\rho)\in L^1(\R^3)$ and $L=0$. More precisely, the Bloch-Floquet decomposition of the Coulomb kernel reads
$$
(|\cdot|)_\bq (\br) = \frac{4\pi}{|\Gamma|} \left( \frac{1}{|\bq|^2} + \sum_{\bK \in \cR^\ast \setminus \left\{0\right\}} \frac{\mathrm{e}^{i\bK \cdot\br}}{|\bq+\bK|^2} \right),
$$
and only the singular component $\frac{4\pi}{|\Gamma|\, |\bq|^2}$, which originates from the long-range of the Coulomb potential, gives a nonzero contribution to $L$.

\medskip

We can deduce from (\ref{eq:rhonu}) and Proposition~\ref{lem:limL} that, in general, the minimizer $Q_{\mnu,\EF}$ to (\ref{eq:model_defaut}) is not trace-class and that the density $\rho_{\mnu,\EF}$ is not an integrable function if the host crystal is anisotropic. Let us detail this point.

Consider some $\mnu \in L^1(\R^3) \cap L^2(\R^3)$ such that $\int_{\R^3} \mnu \neq 0$ and $\|\mnu \star |\cdot|^{-1}\|_{L^2+\cC'} < \beta$. In view of (\ref{eq:boundab}) and Proposition~\ref{prop:chi0}, it holds 
\begin{equation} \label{eq:tr0Q}
\tr_0(Q_{\mnu,\EF})=\tr_0(Q_{1,-V_\mnu}+\widetilde Q_{2,-V_\mnu})=0.
\end{equation}
Assume that $\rho_{\mnu,\EF}$ is in $L^1(\R^3)$. Then a technical lemma (see~\cite[Lemma~4]{CanLew10}) shows that the Fourier transform of the density $\rho_{\widetilde Q_{2,-V_\mnu}}$, corresponding to the nonlinear response terms, is continuous and vanishes at zero. This means that, although it is not known whether $\rho_{\widetilde Q_{2,-V_\mnu}}$ is in $L^1(\R^3)$, this density of charge behaves in the Fourier space as if it was integrable with an integral equal to zero. It follows from (\ref{eq:rhonu}) and Proposition~\ref{prop:chi0} that for each $\sigma \in S^2$,
\begin{equation} \label{eq:Fourier_rho}
\widehat \rho_{\mnu,\EF}(0) =  \lim_{\eta \to 0^+} \widehat{\cL(\rho_{\mnu,\EF}-\mnu)}(\eta\sigma) = (\sigma^T L \sigma) (\widehat \rho_{\mnu,\EF}(0)-\widehat\mnu(0)).
\end{equation}
As by assumption $\widehat\mnu(0)\neq 0$ (since $\int_{\R^3}\mnu \neq 0$), we reach a contradiction unless the matrix $L$ is proportional to the identity matrix. Defining here an isotropic crystal as a crystal for which $L \neq L_0 1$, this proves that, in general, $\rho_{\mnu,\EF}$ is not an integrable function for anisotropic crystals (and this {\it a fortiori} implies that $Q_{\mnu,\EF}$ is not trace-class). 

\medskip

Let us now consider an isotropic crystal. If $Q_{\mnu,\EF}$ were trace-class, then $\rho_{\mnu,\EF}$ would be in $L^1(\R^3)$, and we would deduce from (\ref{eq:tr0Q}) that 
$$
(2\pi)^{3/2} \widehat \rho_{\mnu,\EF}(0) = \int_{\R^3} \rho_{\mnu,\EF} = \tr(Q_{\mnu,\EF})=\tr_0(Q_{\mnu,\EF})=0.
$$
Again, except in the very special case when $L = 1$, this contradicts (\ref{eq:Fourier_rho}) since $\widehat \mnu \neq 0$ by assumption. Thus, in general, $Q_{\mnu,\EF}$ is not trace-class, even for isotropic crystals. We do not know whether the electronic density $\rho_{\mnu,\EF}$ generated by some $\mnu \in L^1(\R^3) \cap L^2(\R^3)$ (this assumption implies $\mnu \in L^{6/5}(\R^3) \hookrightarrow \cC$) in an isotropic crystal is integrable or not. If it is, it follows from (\ref{eq:Fourier_rho}) that, still under the assumption that $\|\mnu \star |\cdot|^{-1}\|_{L^2+\cC'} < \beta$,
$$
\int_{\R^3} \mnu - \int_{\R^3} \rho_{\mnu,\EF} = \frac{\int_{\R^3}\mnu}{1+L_0}.
$$ 
This quantity can be interpreted as the renormalized charge of the defect, which differs from the bare charge $\int_{\R^3}\mnu - \tr_0(Q_{\mnu,\EF})= \int_{\R^3}\mnu$ by a screening factor $\frac{1}{1+L_0}$. This is formally similar to the charge renormalization phenomenon observed in QED (see \cite{GraLewSer-09} for a mathematical analysis).

\subsection{Dielectric operator and macroscopic dielectric permittivity}
\label{sec:macro}

In this section, we focus again on the total potential
\begin{equation} \label{eq:tot_pot}
V_\mnu = (\mnu-\rho_{\mnu,\epsilon_{\rm F}^0}) \star |\cdot|^{-1}
\end{equation}
generated by the total charge of the defect, but we study it in a certain macroscopic limit.

For this purpose, we fix some $\mnu \in L^1(\R^3) \cap L^2(\R^3)$ and introduce for all $\eta > 0$ the rescaled density 
$$
\mnu_\eta(\br) := \eta^3 \mnu(\eta \br).
$$
We then denote by $V_\mnu^\eta$ the total potential generated by $\mnu_\eta$ and the corresponding electronic polarization, \textit{i.e.} 
\begin{equation} \label{eq:Vnueta}
V_\mnu^\eta := (\mnu_\eta-\rho_{\mnu_\eta,\epsilon_{\rm F}^0})\star |\cdot|^{-1},
\end{equation}
and define the rescaled potential
\begin{equation} \label{eq:Wnueta}
W_\mnu^\eta(\br) := \eta^{-1}  \, V_\mnu^\eta \left( \eta^{-1} \br  \right).
\end{equation}
The scaling parameters have been chosen in a way such that in
the absence of dielectric response (\textit{i.e.} for $\cL=0$ and $\widetilde \rho_{Q_{2,-V_\mnu^\eta}}=0$), it holds $W_\mnu^\eta = v_{\rm c}(\mnu) = \mnu \star |\cdot|^{-1}$ for all $\eta > 0$. To obtain a macroscopic limit, we let $\eta$ go to zero.

As $\|(\mnu_\eta\star|\cdot|^{-1})\|_{\cC'} =\|\mnu_{\eta}\|_\cC = \eta^{1/2} \|\mnu\|_{\cC}$, we can apply the results of the previous sections as soon as $\eta$ is small enough. Introducing the family of scaling operators $(U_\eta)_{\eta > 0}$ defined by $(U_\eta f)(\br) = \eta^{3/2} f(\eta\br)$  (each $U_\eta$ is a bijective isometry of $L^2(\R^3)$), the equation linking the density of charge $\mnu$ to the rescaled potential $W^\eta_\mnu$ reads
\begin{equation} \label{eq:scaling}
W^\eta_\mnu = v_{\rm c}^{1/2} U_\eta^\ast \widetilde \epsilon^{-1} U_\eta v_{\rm c}^{1/2} \mnu + \widetilde w^\eta_\mnu,
\end{equation}
where the nonlinear contribution $\widetilde w^\eta_\mnu$ is such that there exists $C \in \R_+$ such that for $\eta$ small enough, $\|\widetilde w^\eta_\mnu\|_{\cC'} \le C\eta$. The macroscopic limit of $W^\eta_\mnu$ therefore is governed by the linear response term, and is obtained as the limit when $\eta$ goes to zero of the family $(U_\eta^\ast \widetilde \epsilon^{-1}U_\eta)_{\eta > 0}$ of bounded self-adjoint operators on $L^2(\R^3)$. 

\medskip

If $\widetilde\epsilon^{-1}$ was translation invariant, that is, if it was commuting with all the translations $\tau_\bR$ for $\bR \in \R^3$, it would be a multiplication operator in the Fourier space (\textit{i.e.} such that for all $f \in L^2(\R^3)$, $\widehat{(\widetilde\epsilon^{-1}f)}(\bk) = \bar\varepsilon^{-1}(\bk) \widehat f(\bk)$ for some function $\R^3 \ni \bk \mapsto \bar\varepsilon^{-1}(\bk) \in \C$). Using the fact that the operator $v_{\rm c}^{1/2}$ is the multiplication operator by $(4\pi)^{1/2}/|\bk|$ in the Fourier space, we would obtain in the limit 
$$
\lim_{\eta \to 0^+} \left( \frac{|\bk|^2}{\bar\varepsilon^{-1}(\eta \bk)}\right)  \widehat W_\mnu(\bk) = 4\pi \widehat \mnu(\bk).
$$ 
As the operator $\widetilde\epsilon^{-1}$ actually commutes only with the translations of the lattice $\cR$, the above argument cannot be applied. On the other hand, it can be proved, using Bloch-Floquet decomposition, that $W^\eta_\mnu$ has a limit $W_\mnu$ when $\eta$ goes to zero, and that this limits satisfies
\begin{equation} \label{eq:limW}
\lim_{\eta \to 0^+} \left( \frac{|\bk|^2}{[\widetilde\epsilon^{-1}]_{00}(\eta \bk)}\right)  \widehat W_\mnu(\bk) = 4\pi \widehat \mnu(\bk),
\end{equation}
where $[\widetilde\epsilon^{-1}]_{00}(\bq)$ is the entry of the Bloch matrix of the $\cR$-periodic operator $\widetilde\epsilon^{-1}$ corresponding to $\bK=\bK'=0$. Besides,
\begin{equation} \label{eq:defEM}
\lim_{\eta \to 0^+} \left( \frac{|\bk|^2}{[\widetilde\epsilon^{-1}]_{00}(\eta \bk)}\right) = \bk^T \epsilon_{\rm M} \bk,
\end{equation}
where $\epsilon_{\rm M}$ is a $3 \times 3$ symmetric, positive definite matrix. Transforming back (\ref{eq:limW}) in the physical space, we obtain the macroscopic Poisson equation (\ref{eq:macroscopicPoisson2}). Let us formalize this central result in a theorem.

\medskip

\begin{theorem}\label{thm:homogenization}
There exists a $3 \times 3$ symmetric matrix $\epsilon_{\rm M} \ge 1$ such
that for all $\mnu \in L^1(\R^3) \cap L^2(\R^3)$, the rescaled potential 
$W_\mnu^\eta$ defined by (\ref{eq:Wnueta}) converges to $W_\mnu$ weakly in $\cC'$ when $\eta$ goes to zero, where $W_\mnu$ is the unique
solution in $\cC'$ to the elliptic equation
$$
-\div(\epsilon_{\rm M} \nabla W_\mnu ) = 4\pi\mnu.
$$
The matrix $\epsilon_{\rm M}$ is proportional to the identity matrix if the
host crystal has the symmetry of the cube.
\end{theorem}

From a physical viewpoint, the matrix $\epsilon_{\rm M}$ is the
electronic contribution to the macroscopic dielectric tensor of 
the host crystal. Note that the other contribution, originating from the displacements of the nuclei~\cite{PicCohMar-70}, is not taken into account in this study. 

The matrix $\epsilon_{\rm M}$ can be computed from the Bloch-Floquet decomposition of $H^0_{\rm per}$ as follows. The operator $\widetilde \epsilon = v_{\rm c}^{-1/2} \epsilon v_{\rm c}^{1/2}$ being $\cR$-periodic, it can be represented by the Bloch matrices $([\widetilde \epsilon_{\bK \bK'}(\bq)]_{\bK,\bK' \in \cR^\ast})_{\bq \in \Gamma^\ast}$. It is proven in~\cite{CanLew10} that each entry of the Bloch matrix $\widetilde{\epsilon}_{\bK,\bK'}(\eta\sigma)$ has a limit when $\eta$ goes to $0^+$ for all fixed $\sigma\in S^2$. Indeed, 
$$
\lim_{\eta\to0^+}\widetilde{\epsilon}_{0,0}(\eta\sigma)=1+\sigma^T L \sigma
$$
where $L$ is the $3 \times 3$ non-negative symmetric matrix defined
in~\eqref{eq:mk}.  
When $\bK,\bK'\neq0$, $\widetilde{\epsilon}_{\bK,\bK'}(\eta\sigma)$ has a limit at
$\eta=0$, which is independent of $\sigma$ and which we simply denote by
$\widetilde{\epsilon}_{\bK,\bK'}(0)$. When $\bK=0$ but $\bK'\neq0$, the limit is a
linear function of $\sigma$: for all $\bK' \in \cR^\ast \setminus
\left\{0\right\}$,  
$$
\lim_{\eta\to0^+} \widetilde{\epsilon}_{0,\bK'}(\eta\sigma)=\beta_{\bK'} \cdot \sigma,
$$
for some $\beta_{\bK'} \in \C^3$. Both $\widetilde\epsilon_{\bK\bK'}(0)$ ($\bK,\bK'\neq 0$) and $\beta_\bK$ can be computed from the eigenvalues $\epsilon_{n,\bq}$ and eigenvectors $u_{n,\bq}$ of the Bloch-Floquet decomposition of $\Hper$ by formulae similar to (\ref{eq:mk}).
As already mentioned, the electronic contribution to the macroscopic
dielectric permittivity is the $3 \times 3$ symmetric tensor
defined as \cite{BarRes-86}
\begin{equation} \label{eq:epsilonM}
\forall \bk \in \R^3, \quad 
\bk^T \epsilon_{\rm M} \bk = \lim_{\eta \to 0^+} \frac{|\bk|^2}{[
  \widetilde\epsilon^{-1}]_{00}(\eta \bk)}.
\end{equation}  
By the Schur complement formula, it holds
$$ 
\frac{1}{[\widetilde\epsilon^{-1}]_{00}(\eta \bk)}=\widetilde\epsilon_{00}(\eta
\bk)-\sum_{\bK,\bK'\neq0}{\widetilde{\epsilon}_{0,\bK}(\eta \bk)}[C(\eta
\bk)^{-1}]_{\bK,\bK'}\widetilde{\epsilon}_{\bK',0}(\eta \bk) 
$$
where $C(\eta \bk)^{-1}$ is the inverse of the matrix
$C(\eta \bk)=[\widetilde\epsilon_{\bK\bK'}(\eta \bk)]_{\bK,\bK'\in \cR^\ast\setminus\left\{0\right\}}$. 
This leads to
$$ 
\lim_{\eta\to0^+}\frac{|\bk|^2}{[\widetilde\epsilon^{-1}]_{00}(\eta \bk)}
=|\bk|^2+\bk^TL\bk-\sum_{\bK,\bK' \in \cR^\ast\setminus\left\{0\right\}} (\beta_{\bK}
\cdot \bk) [C(0)^{-1}]_{\bK,\bK'}  
\overline{(\beta_{\bK'} \cdot \bk)}
$$
where $C(0)^{-1}$ is the inverse of the matrix
$C(0)=[\widetilde\epsilon_{\bK\bK'}(0)]_{\bK,\bK'\in
  \cR^\ast\setminus\left\{0\right\}}$. Therefore,
\begin{equation} \label{eq:epsilon_M}
\epsilon_{\rm M} = 1 + L -\sum_{\bK,\bK'\in
  \cR^\ast\setminus\left\{0\right\}} \beta_{\bK} [C(0)^{-1}]_{\bK,\bK'}
\beta_{\bK'}^\ast.
\end{equation}
As already noticed in \cite{BarRes-86}, it holds
$$
1 \leq \epsilon_{\rm M} \leq 1 + L.
$$

Formula~\eqref{eq:epsilon_M} has been used in numerical simulations for estimating the macroscopic dielectric permittivity of real insulators and
semiconductors~\cite{BarRes-86,HybLou-87a,HybLou-87b,EngFar-92,GajHumKreFurBec-06}.
Direct methods for evaluating $\epsilon_{\rm M}$, bypassing the
inversion of the matrix $C(0)$, have also been
proposed~\cite{ResBal-81,KunTos-84}.  

\subsection{Time-dependent response}
\label{sec:TDexpansion}

We study in this section the variation of the electronic state of the crystal when the mean-field Hamiltonian $H^0_{\rm per}$ of the perfect crystal is perturbed by a time-dependent effective potential $v(t,\br)$ of the form (\ref{eq:TD_pot}).
The mathematical proofs of the results announced in this section will be given in \cite{CanSto}.

\medskip

Let 
$$
H_v(t)=H^0_{\rm per}+v(t,\cdot)=-\frac 12 \Delta + V_{\rm per}+v(t,\cdot). 
$$
Under the assumption that $\rho^{\rm nuc}_{\rm per} \in L^2_{\rm per}(\Gamma)$ (smeared nuclei), the mean-field potential $V_{\rm per}$ is $\cR$-periodic and in $C^0(\R^3) \cap L^\infty(\R^3)$. 
Besides, there exists a constant $C > 0$ such that $\|\rho \star |\cdot|^{-1}\|_{L^\infty} \le C \|\rho\|_{L^2\cap \cC}$ for all $\rho \in L^2(\R^3) \cap \cC$, so that the time-dependent perturbation~$v$ belongs to $L^1_{\rm loc}(\R,L^\infty(\R^3))$.

Let us now define the propagator $(U_v(t,s))_{(s,t) \in \R \times \R}$ associated with the 
time-dependent Hamiltonian $H_v(t)$ following~\cite[Section~X.12]{ReeSim2}.
To this end, consider first the propagator $U_0(t) = \mathrm{e}^{-itH^0_{\rm per}}$ associated with the time-independent Hamiltonian $H^0_{\rm per}$, and the perturbation in the so-called interaction picture:
\[
v_{\rm int}(t) = U_0(t)^* v(t) U_0(t).
\]
Standard techniques (see for instance~\cite[Section~5.1]{Pazy})
allow to show the existence and uniqueness of the family of unitary propagators 
$(U_{\rm int}(t,s))_{(s,t) \in \R \times \R}$ 
associated with the bounded operators~$(v_{\rm int}(t))_{t \in \R}$,
with
\[
U_{\rm int}(t,t_0) = 1 - \ri \int_{t_0}^t v_{\rm int}(s) U_{\rm int}(s,t_0) \, ds.
\]
Therefore, $U_v(t,s) = U_0(t) U_{\rm int}(t,s) U_0(s)^*$ satisfies the integral equation
\begin{equation} \label{eq:IEUv}
U_v(t,t_0) = U_0(t-t_0) - \ri \int_{t_0}^t U_0(t-s) v(s) U_v(s,t_0) \, ds.
\end{equation}

Denoting by $\gamma^0$ the density operator of the crystal at time $t=0$, the dynamics of the system is governed by the evolution equation
\begin{equation} \label{eq:time_evolution}
\gamma(t) = U_v(t,0) \gamma^0 U_v(t,0)^*.
\end{equation}
Note that the conditions $\gamma^0 \in {\cal S}(L^2(\R^3))$ and $0 \le \gamma^0 \le 1$ are automatically propagated by (\ref{eq:time_evolution}).

\medskip

Considering $v(t)$ as a perturbation of the time-independent Hamiltonian $H^0_{\rm per}$, and $\gamma(t)$ as a perturbation of the ground state density operator $\Gper$, it is natural to follow the same strategy as in the time-independent setting and introduce
$$
Q(t) = \gamma(t)-\Gper.
$$
Using (\ref{eq:IEUv}), (\ref{eq:time_evolution}), and the fact that $\Gper$ is a steady state of the system in the absence of perturbation ($U_0(t)\Gper U_0(t)^\ast=\Gper$), an easy calculation shows that $Q(t)$ satisfies the integral equation
\begin{equation} \label{eq:time_evolution_Q}
Q(t)= U_0(t)Q(0)U_0(t)^\ast - \ri \int_0^t U_0(t-s) [v(s),\Gper+Q(s)] U_0(t-s)^\ast \, ds.
\end{equation}

We now assume that $\gamma^0=\Gper$, \textit{i.e.} $Q(0)=0$, and write (formally for the moment) $Q(t)$ as the series expansion 
\begin{equation}
Q(t) = \sum_{n=1}^{+\infty} Q_{n,v}(t),\label{eq:TDexpansion}
\end{equation}
where the operators $Q_{n,v}(t)$ are obtained, as in the time-independent case, by identifying terms involving $n$ occurrences of the potential~$v$. In particular, the linear response is given by
\begin{equation}
  \label{eq:Q1v}
  Q_{1,v}(t) = - \ri   \int_0^t U_0(t-s) \left [ v(s), \Gper \right]  U_0(t-s)^*  \, ds,
\end{equation}
and the following recursion relation holds true
\begin{equation}
  \label{eq:recursionQnv}
\forall n \ge 2, \quad  Q_{n,v}(t) = - \ri   \int_0^t U_0(t-s) \left [ v(s), Q_{n-1,v}(s) \right]  U_0(t-s)^*  \, ds.
\end{equation}

\medskip

It is proved in \cite{CanSto} that for any $n \geq 1$ and any $t \geq 0$, the operator $Q_{n,v}(t)$ in~\eqref{eq:TDexpansion} belongs to $\cQ$ and satisfies 
$$
\forall \psi \in L^2(\R^3), \quad \langle \psi|Q_{n,v}(t)|\psi \rangle_{L^2} = 0.
$$
In particular, $\tr_0(Q_{n,v}(t)) = 0$. Besides, there exists $b \in \R_+$ such that for all $t \ge 0$
  \[
  \| Q_{n,v}(t) \|_{\mathcal{Q}} \leq b^n \int_0^t \int_0^{t_1} \dots \int_0^{t_{n-1}} 
  \| \rho(t_1) \|_{L^2 \cap \cC} \dots \| \rho(t_n) \|_{L^2 \cap \cC} \, dt_n \dots dt_1,
  \]
and there exists $T > 0$ such that the series expansion (\ref{eq:TDexpansion}) converges in ${\cal Q}$ uniformly on any compact subset of $[0,T)$. Lastly, $T=+\infty$ if $\rho \in  L^\infty(\R_+,L^2(\mathbb{R}^3) \cap \cC)$.

\medskip

As in the time-independent setting, the frequency-dependent dielectric properties of the crystal can be obtained from the linear response (\ref{eq:Q1v}), by defining the time-dependent independent-particle polarization operator
\begin{equation}
\label{eq:chi0}
\begin{array}{rcl}
\chi_0 \ : \ L^1(\R,v_{\rm c}(L^2(\R^3)\cap\cC)) & \to &  L^\infty(\R,L^2(\R^3)\cap\cC)\\
v & \mapsto & \rho_{Q_{1,v}}
\end{array}
\end{equation}
and the time-dependent operators ${\cal L}=-\chi_0v_{\rm c}$, $\epsilon=v_{\rm c}(1+\cL)v_{\rm c}^{-1}$, $\epsilon^{-1}=v_{\rm c}(1+\cL)^{-1}v_{\rm c}^{-1}$, and $\widetilde\epsilon= v_{\rm c}^{-1/2}\epsilon v_{\rm c}^{1/2}$. Due to the invariance of the linear response with respect to translation in time, all these operators are convolutions in time. In addition they are $\cR$-periodic in space. They can therefore be represented by frequency-dependent Bloch matrices $[T_{\bK,\bK'}(\omega,\bq)]$, with $\bK$, $\bK'$ in $\cR^\ast$, $q \in \Gamma^\ast$ and $\omega \in \R$. The Adler-Wiser formula states that the (electronic contribution of the) frequency-dependent macroscopic dielectric permittivity is given by the formula
$$
\forall \bk \in \R^3, \quad \bk^T \cFt\epsilon_{\rm M}(\omega) \bk = \lim_{\eta \to 0^+} \left( \frac{|\bk|^2}{[\widetilde\epsilon^{-1}]_{00}(\omega,\eta \bk)}\right) .
$$
The mathematical study of this formula and of its possible derivation from rigorous homogenization arguments, is work in progress.

\medskip

We finally consider the self-consistent Hartree dynamics defined by
\begin{equation}
  \label{NL_dynamics_integral}
  Q(t) = U_0(t) Q^0 U_0(t)^* - \ri \int_0^t U_0(t-s) \Big[ v(s) + v_{\rm c}(\rho_{Q(s)}),\Gper + Q(s) \Big ] U_0(t-s)^* ds,
\end{equation}
for an initial condition $Q^0 \in \cK$, and for an external potential $v(t)=v_{\rm c}(\mnu(t))$, where $\mnu(t) \in L^2(\R^3) \cap {\cal C}$ for all $t$. The solution $Q(t)$ of~\eqref{NL_dynamics_integral} is such that $\gamma(t)=\Gper+Q(t)$ satisfies, formally, the time-dependent Hartree equation
$$
i \frac{d\gamma}{dt}(t) = \left[-\frac 12 \Delta +(\rho_{\gamma(t)}- \rho^{\rm nuc}_{\rm per}-\mnu(t)) \star |\cdot|^{-1}, \gamma(t) \right].
$$

\medskip

The following result \cite{CanSto} shows the well-posedness of the nonlinear Hartree dynamics.

\medskip

\begin{theorem} \label{Th:TDrHF}
Let $\mnu \in C^1(\mathbb{R}_+,L^2(\R^3) \cap \cC)$. Then, for any $Q^0 \in \cK$, the time-dependent Hartree equation~\eqref{NL_dynamics_integral} has a unique solution in $C^0(\R_+,\cQ)$. Besides, for all $t \ge 0$, $Q(t) \in \cK$ and $\tr_0(Q(t)) = \tr_0(Q^0)$.
\end{theorem}

\section*{Appendix: trace-class and self-adjoint operators}

It is well-known that any compact self-adjoint operator $A$ on a separable Hilbert space ${\cal H}$ can be diagonalized in an orthonormal basis set:
\begin{equation} \label{eq:dec_CSA}
A = \sum_{i=1}^{+\infty} \lambda_i \, |\phi_i\rangle \, \langle \phi_i|,
\end{equation}
where $\langle \phi_i|\phi_j \rangle = \delta_{ij}$, and where the sequence $(\lambda_i)_{i \ge 1}$ of the (real) eigenvalues of $A$, counted with their multiplicities, converges to zero. We have formulated (\ref{eq:dec_CSA}) using again Dirac's bra-ket notation. The conventional mathematical formulation for (\ref{eq:dec_CSA}) reads
$$
\forall \phi \in {\cal H}, \quad A\phi = \sum_{i=1}^{+\infty} \lambda_i \, \langle \phi_i|\phi \rangle \, \phi_i. 
$$
A compact self-adjoint operator $A$ is called trace-class if 
$$
\sum_{i=1}^{+\infty} |\lambda_i|<\infty.
$$
The trace of $A$ is then defined as
$$
\tr(A) := \sum_{i=1}^{+\infty} \lambda_i = \sum_{i=1}^{+\infty} \langle e_i|A|e_i\rangle,
$$
the right-hand side being independent of the choice of the orthonormal basis $(e_i)_{i \ge 1}$. Note that if $A$ is a non-negative self-adjoint operator, the sum
$\sum_{i=1}^{+\infty} \langle e_i|A|e_i\rangle$ makes sense in $\R_+ \cup \left\{+\infty\right\}$ and its values is independent of the choice of the orthonormal basis $(e_i)_{i \ge 1}$. We can therefore give a sense to $\tr(A)$ for {\em any} non-negative self-adjoint operator $A$, and this number is finite if and only if $A$ is trace-class.

The notion of trace-class operators can be extended to non-self-adjoint operators \cite{ReeSim4,Simon-79}, but we do not need to consider this generalization here.

\medskip

By definition, a compact operator $A$ is Hilbert-Schmidt if $A^\ast A$ is trace-class.  A compact self-adjoint operator $A$ on ${\cal H}$ decomposed according to (\ref{eq:dec_CSA}) is Hilbert-Schmidt if and only if 
$$
\sum_{i\geq1}|\lambda_i|^2 <\infty.
$$
Obviously any trace-class self-adjoint operator is Hilbert-Schmidt, but the converse is not true. 

In this contribution, we respectively denote by $\gS_1$ and $\gS_2$ the spaces of trace-class and Hilbert-Schmidt operators acting on $L^2(\R^3)$. We also denote by ${\cal S}(L^2(\R^3))$ the vector space of the bounded self-adjoint operators on $L^2(\R^3)$.

A classical result states that if $A$ is a Hilbert-Schmidt operator on $L^2(\R^3)$, then it is an integral operator with kernel in $L^2(\R^3 \times \R^3)$. This means that there exists a unique function in $L^2(\R^3 \times \R^3)$, also denoted by $A$ for convenience, such that
\begin{equation} \label{eq:kernel}
\forall \phi \in L^2(\R^3), \quad (A\phi)(\br) = \int_{\R^3} A(\br,\br') \, \phi(\br') \, d\br'.
\end{equation}
Conversely, if $A$ is an operator on $L^2(\R^3)$ for which there exists a function $A \in L^2(\R^3 \times \R^3)$ such that (\ref{eq:kernel}) holds, then $A$ is Hilbert-Schmidt.

If $A$ is a self-adjoint Hilbert-Schmidt operator on $L^2(\R^3)$ decomposed according to (\ref{eq:dec_CSA}), then its kernel is given by
$$
A(\br,\br')=\sum_{i\geq1}\lambda_i\,\phi_i(\br) \phi_i(\br').
$$
If, in addition $A$ is trace-class, then the density $\rho_A$, defined as
$$
\rho_A(\br)  = \sum_{i=1}^{+\infty} \lambda_i |\phi_i(\br)|^2,
$$
is a function of $L^1(\R^3)$ and it holds
$$
\tr(A) = \sum_{i=1}^{+\infty} \lambda_i = \int_{\R^3} \rho_A(\br)\, d\br.
$$
For convenience, we use the abuse of notation which consists in writing 
$\rho_A(\br) = A(\br,\br)$ even when the kernel of $A$ is not continuous on
the diagonal $\{\br=\br'\}\subset\R^6$.

\end{document}